\documentclass[10pt,a4paper]{article}
\usepackage{graphicx,amsfonts,amssymb,amsmath, hyperref}
\usepackage{a4wide}

\newif\ifhyper
\hypertrue
\ifhyper
\hypersetup{
   citecolor = {green},
   colorlinks = {true}, 
   urlcolor = {blue} 
}
\fi
\newcommand{\beq}{\begin{equation}}
\newcommand{\eeq}{\end{equation}}
\newcommand{\beqa}{\begin{eqnarray}}
\newcommand{\eeqa}{\end{eqnarray}}
\newcommand{\ket} [1] {\vert #1 \rangle}

\begin{document}

\title{Advances on Tensor Network Theory: \\ Symmetries, Fermions, Entanglement, and Holography}

\author{Rom\'an Or\'us \footnote{\emph{E-mail address:} roman.orus@uni-mainz.de} \\ 
\multicolumn{1}{p{0.9\textwidth}}{\centering\emph{Institute of Physics, Johannes Gutenberg University, 55099 Mainz, Germany}}}
\maketitle

\begin{abstract}

This is a short review on selected theory developments on Tensor Network (TN) states for strongly correlated systems. Specifically, we briefly review the effect of symmetries in TN states, fermionic TNs, the calculation of entanglement Hamiltonians from Projected Entangled Pair States (PEPS), and the relation between the Multi-scale Entanglement Renormalization Ansatz (MERA) and the AdS/CFT or gauge/gravity duality. We stress the role played by entanglement in the emergence of several physical properties and objects through the TN language. Some recent results along these lines are also discussed. 

\end{abstract}

\newpage 
\enlargethispage{1cm}
\tableofcontents
\clearpage

\section{Introduction}

Tensor Network (TN) states are a new language, based on entanglement, for quantum many-body states \cite{TN}. TNs are well-known in the field of strongly-correlated systems, since they are the core of the so-called TN numerical simulation methods. Such methods have extended the ideas of the Density Matrix Renormalization Group (DMRG) \cite{dmrg}, originally designed to tackle $1d$ quantum lattice systems, to a wide variety of other physical situations. These methods offer nowadays a promising alternative to other numerical approaches. But independently of numerics, recent developments on the theory of TN states have also been remarkable, with direct implications for strongly correlated systems and for other fields of physics as well. 

This colloquium paper briefly reviews four of these theory developments. To be specific, we consider (i) the effect of symmetries in TN states and the emergence of spin networks, (ii) fermionic TNs and the relation between fermionic statistics and graphical projections, (iii) the emergence of entanglement Hamiltonians from Projected Entangled Pair States (PEPS) via holography, and (iv) the parallelism between the Multi-scale Entanglement Renormalization Ansatz (MERA) and the AdS/CFT or gauge/gravity duality in quantum gravity. We will discuss on the direct implications of these results both in the theory of TNs, as well as in the numerical simulation of strongly-correlated systems. 

The text is organised as follows: in Sec.2 we explain some theory basics and comment on the main families of TN states and possible classifications. In Sec.3 we briefly review the effect of symmetries in TNs. Sec.4 deals with fermionic TNs, showing how  second quantisation leads to a consistent picture of fermionic statistics based on graphical projections. In Sec.5 we consider the calculation of entanglement spectrum and entanglement Hamiltonians of a PEPS, which offers a wonderful example of the holographic principle at work. Sec.6 deals briefly with the parallelism between the MERA and the AdS/CFT duality. Finally, in Sec.7 we offer some concluding remarks. For those interested in more details or further reading, proper references to each topic are given within each section. 

\section{Some formalities}

To begin with, we remind some of the basics on TN theory. We focus on reviewing some of the most important families of TN states such as Matrix Product States (MPS), PEPS and MERA, together with some of their basic properties. This has certainly been discussed already in the past in many references (see, e.g., \cite{TN} and references therein). However, we wish to refresh them here a little bit in order understand better the developments to be reviewed later, as well as to fix the notation. 

\subsection{Tensor networks, or the breakdown of the quantum state} 

Tensor networks are representations of quantum many-body states of matter based on their local entanglement structure \cite{TN}. In a way, we could say that one uses entanglement to build up the many-body wave function. To see how this is possible, consider, e.g., a quantum many-body system of $N$ spins-$1/2$, 
\beq
\ket{\psi} = \sum_{i_1 i_2 \cdots i_N}C_{i_1 i_2 \cdots i_N} \ket{i_1 i_2 \cdots i_N}. 
\eeq
Any wave function of the system can be described, up to normalisation and in a inefficient way, in terms of the $2^N$ complex coefficients $C_{i_1 i_2 \cdots i_N}$ once an individual basis $\ket{i_r}$ for spins $r = 1,2, \ldots , N$ is chosen. Such coefficients can be understood as forming a tensor $C$ with $N$ indices, where each one of the indices can take two values (say, spin``up" and ``down"). 

\begin{figure}
\begin{centering}
\includegraphics[width=8cm]{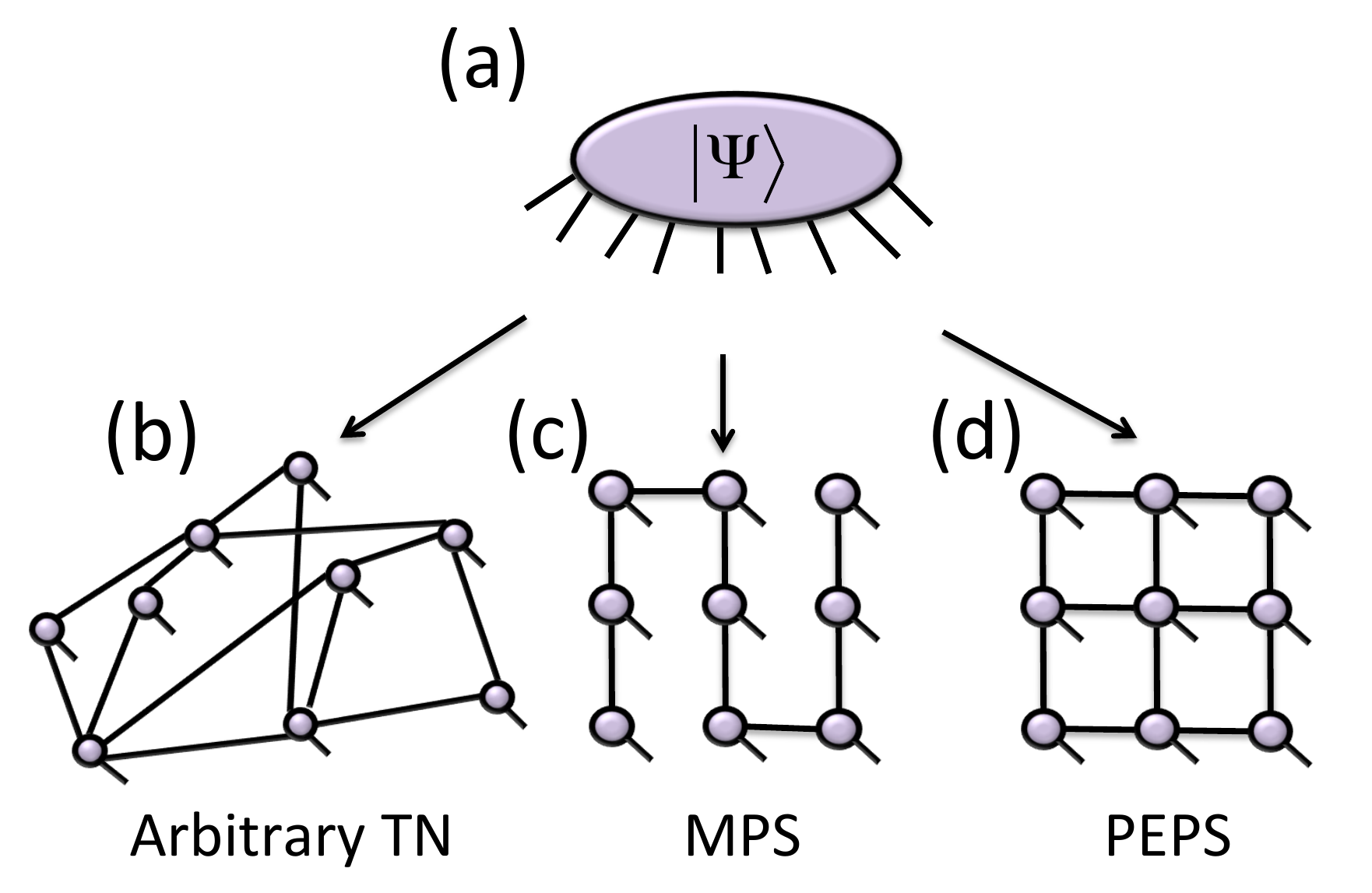} 
\par\end{centering}
\caption{(color online) (a) a Quantum state $\ket{\Psi}$ for $9$ local degrees of freedom (e.g., spins) is decomposed as (b) an arbitrary TN, (b) a $1d$ MPS, and (c) a $2d$ PEPS.} \label{Fig1}
\end{figure}

To reduce the complexity in the representation of the many-body state, one can replace tensor $C$ by a network of smaller interconnected tensors, where each one of the tensors has a small number of coefficients, see Fig.\ref{Fig1} for an example. In this figure we have used a diagrammatic notation: tensors are represented by shapes, and indices by lines. Contracted indices, i.e., common indices between tensors over which there is a sum for all their possible values, are represented by lines going from one shape to another, i.e., indices connecting different tensors. Further details on the diagrammatic representation of tensor networks can be found, for instance, in Ref.\cite{TN}. 

As explained in Ref.\cite{TN}, we can think of this approach as decomposing the quantum state in ``fundamental DNA blocks'', namely, tensors with less parameters \footnote{One may even think that \emph{the wave function itself is quantised} in terms of a discrete set of smaller tensors. Yet, notice that this notion of quantisation has nothing to do with promoting the wave-function to an operator -as in second quantisation-, but rather with splitting it into a set of ``fundamental units" with each choice of TN being a different (non-observable) ``quantisation scheme". In any case, this is purely an analogy and of course it is just a matter of taste.}. Reversely, the small tensors are the DNA of the wave function, in the sense that all the properties of the many-body quantum state can be read from the individual small tensors alone. Such a construction is called a tensor network, and it depends typically on $O({\rm poly}(N))$ parameters (since the number of tensors is indeed polynomial), thus being a computationally efficient description of the quantum state of the many-body system.

The TN construction also involves the appearance of extra degrees of freedom in the system, which ``glue" the different tensors together. These new degrees of freedom are represented by the connecting indices amongst the tensors in the TN, which are called \emph{bond indices}. In practice they provide the structure of the many-body entanglement in the quantum state, and their rank $D$ (that is, the number of different values that each one of these indices can take) is a quantitative measure of entanglement in the quantum state. For instance, $D=1$ corresponds to a separable product state (i.e., mean field theory), while any $D>1$ provides non-trivial entanglement properties. The largest of these ranks in a TN is called the \emph{bond dimension} of the TN. It is also well-known that $D$ measures the degree of entanglement in the system, and can be related to the so-called area-law for the entanglement entropy and generalisations thereof \cite{TN, MERA, braMERA}. Intuitively, one could say that entanglement is the glue between the different patches that quantise the wave-function of a system. 

There are a number of motivations to build TN states. We already discussed that these states offer efficient descriptions of quantum states of matter. But moreover, they can be regarded as a generalisation of mean-field theory (which uses product states, i.e., $D=1$, to describe many-body systems). It is also well known that TN states can be seen in terms of a collection of maximally entangled states projected locally on some Hilbert spaces of smaller dimension \cite{aklt, PEPS2}. But most importantly, TN states are relevant since they codify the correct structure of entanglement in many-body states, in turn targeting the (zero-measure) relevant corner of the Hilbert space for the description of low-energy states of Hamiltonians with local interactions (see, e.g., \cite{TN, Hastings}). As already proven \cite{Hastings, parent}, locality has something to say in the structure of low-energy quantum states of matter: the wave-function is built locally by sewing fundamental patches of the quantum state (i.e., the tensors) using entanglement\footnote{For the sake of this review we shall not discuss the case of Hamiltonians with long-range interactions.}. 

\subsection{Classifying tensor network states}

TN states may be classified from different perspectives. For instance, one could classify them as either (i)  \emph{discrete}, i.e., those for quantum states on lattices, or (ii) \emph{continuous}, i.e., those for quantum states on a continuum. Continuous TN states have been proposed rather recently and are a hot research topic nowadays \cite{cMPS, cMERA, cTN, Ryu, Swingle2}. In this short review, however, we focus on discrete TNs for quantum many-body states on lattices, for which there is a more extensive literature available. 

From a different perspective, TN states may also be classified in terms of the dimensions along which tensors are spanned. According to these criteria, one can talk about TN (i) \emph{without extra dimensions} (such as Matrix Product States (MPS) \cite{dmrg, tebd, itebd} and Projected Entangled Pair States (PEPS) \cite{PEPS2}), and (ii) \emph{with extra dimensions} (such as Tree Tensor Networks (TTN) \cite{ttncan, unttn, isottn} and the Multi-scale Entanglement Renormalization Ansatz (MERA) \cite{MERA}). Extra dimensions in a TN state are usually holographic, in the sense that they define a ``tensor bulk space" such that the physical quantum state is recovered at its boundary. Moreover, such extra dimensions may be interpreted in terms of a renormalization group scale. Let us quickly review some of the main families of TN states, and some of their key properties, according to this classification criteria. 

\begin{figure}
\begin{centering}
\includegraphics[width=10cm]{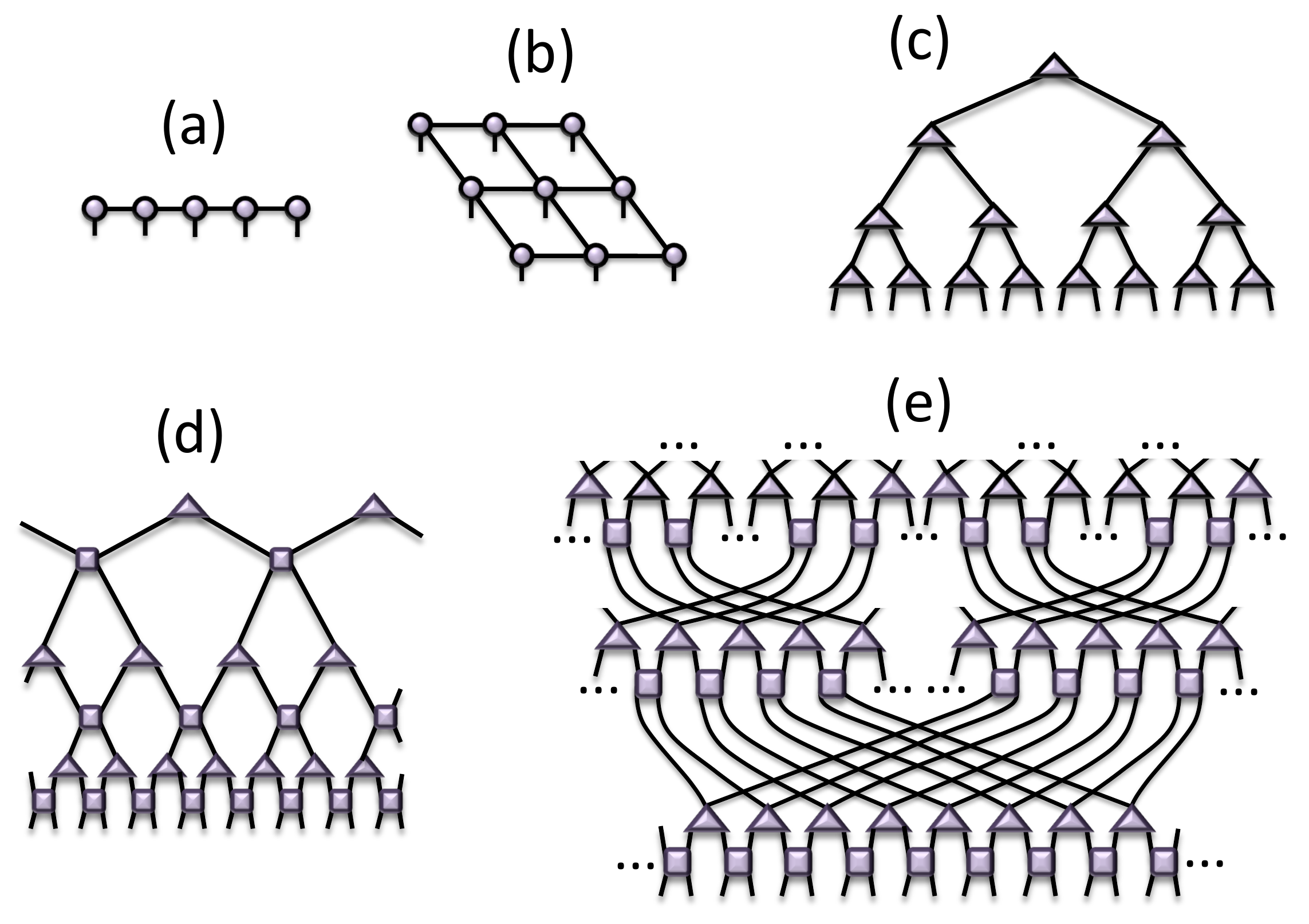} 
\par\end{centering}
\caption{(color online) Examples of TN states: (a) MPS; (b) PEPS; (c) TTN; (d) MERA; (e) branching-MERA.} \label{Fig2}
\end{figure}

\subsubsection{TNs without extra dimensions}

This is the case in which tensors are spanned along the physical directions of the underlying system, i.e., along a $m$ directions for a system in $m$ spatial dimensions. 

\emph{\underline{Matrix Product States (MPS).-}}

The family of MPS is probably the most famous example of TN states. This is mainly because MPS are behind some very powerful methods to simulate $1d$ quantum many-body systems, most prominently the Density Matrix Renormalization Group (DMRG) algorithm \cite{dmrg, pbc, dmrgrev1, dmrgrev2}. But it is also behind other very useful methods such as Time-Evolving Block Decimation (TEBD) \cite{tebd, itebd} and Power Wave Function Renormalization Group (PWFRG) \cite{pwfrg}. MPS have been discussed in the literature already many times, so here we will just remember some of their basic properties, and without entering into too many details.  The interested reader is referred to, e.g., Ref.\cite{TN, dmrgrev1, dmrgrev2}, for further details. 

Loosely speaking, MPS are TN states that correspond to a $1d$ array of tensors\footnote{Mathematicians sometimes call this the \emph{Tensor Train Decomposition} \cite{lowrank}.}. We have one tensor per site in the many-body system. Bond indices can take up to $D$ values, and the open indices correspond to the physical degrees of freedom of the local Hilbert spaces at every lattice site. See Fig.\ref{Fig2}(a) for an example.  

Let us now quickly review -- without entering into proofs -- some of the important properties of MPS. First, by assuming translational invariance one can take the \emph{thermodynamic limit} of an MPS (called inifnite-MPS, or iMPS) \cite{itebd}. This is just achieved by defining a unit cell of tensors that is repeated along the system infinitely-many times. Second, they are \emph{dense}, i.e., by increasing $D$ sufficiently one can always access the entire Hilbert space, yet the value of $D$ would need to be exponentially large in the size of the system to cover all possible states. Third, they obey a \emph{$1d$ version of the area-law for the entanglement entropy of a block}, namely, $S(L) = -{\rm tr}(\rho_L \log \rho_L) \le 2\log D$,  with $\rho_L$ the reduced density matrix of a block of $L$ contiguous sites. Such entanglement scaling is an exact result in ground states of gapped $1d$ Hamiltonians with local interactions and for large $L$ \cite{has}. More precisely, for these systems $S(L) \sim  {\rm constant}$ for $L\gg1$ \cite{arealaw}, and therefore  MPS turn out to reproduce the correct entanglement structure of such states of matter. Fourth, MPS have a \emph{finite correlation length}. For a derivation of the correlation length of an MPS see, e.g, Refs.\cite{dmrgrev1, dmrgrev2, TN}. Fifth, expectation values of local observables can be computed \emph{both exactly and efficiently}. In fact, the exact calculation of the scalar product between two MPS for $N$ sites can always be done exactly in a time $O(N p D^3)$, with $p$ the local Hilbert space dimension (e.g., $p=2$ for a chain of spin-$1/2$ particles) \cite{TN}.  In general, expectation values of local observables such as correlation functions, energies, and local order parameters, can also be computed using the same kind of tensor manipulations as for the norm, and with a similar leading computational cost. Sixth, MPS with open boundary conditions (either for a finite or infinite system) admit the so-called \emph{canonical form}, where bond indices can be associated to orthonormal left/right Schmidt basis, see Refs.\cite{tebd, itebd}. The canonical form turns out to be very useful in a variety of contexts, e.g., in defining local truncation schemes in time-evolution algorithms \cite{tebd, itebd}, as well as in identifying symmetry-protected topological order in $1d$ systems \cite{pollmann}, just to mention a couple of examples. \color{black} For completeness, let us also mention that the existence of a \emph{pseudo-canonical form} for infinite and periodic systems (see first paper in Ref.\cite{TN} and references therein). \color{black}

\emph{\underline{Projected Entangled Pair States (PEPS).-}}

The family of PEPS \cite{PEPS2} is the natural generalisation of MPS to higher spatial dimensions. Here we shall only consider the $2d$ case, see Fig.\ref{Fig2}(b). As before, bond indices in a PEPS are of dimension $D$. $2d$ PEPS are also at the basis of several methods to simulate $2d$ quantum lattice systems, e.g., PEPS \cite{PEPS2} and infinite-PEPS \cite{iPEPS} algorithms, as well as Tensor Renormalization Group (TRG) \cite{TRG}, Second Renormalization Group (SRG) \cite{SRG}, Higher-Order Tensor Renormalization Group (HOTRG) \cite{HOTRG}, and methods based on Corner Transfer Matrices (CTM) and Corner Tensors \cite{dctm,ctmrg,ctens}. These methods have been discussed extensively in the literature, see, e.g., Ref.\cite{TN} and references therein. 

Let us quickly review the main properties of PEPS without proving them. First, they can be defined in the \emph{thermodynamic limit} (as MPS) whenever they can be chosen to be invariant under $2d$ translations. Second, they are also \emph{dense}, so that by increasing $D$ one recovers progressively the whole Hilbert space. But again, $D$ may end up being exponentially large in one of the physical directions of the system in order to be able to cover the entire space. Third, PEPS satisfy the \emph{$2d$ area-law for the entanglement entropy.} Namely, the entanglement entropy of an $L \times L$ block of a PEPS with bond dimension $D$ -- say, on a $2d$ square lattice -- obeys $S(L) \le 4L \log D$. This property is satisfied by many interesting quantum states, such as some ground states and low-energy excited states of $2d$ Hamiltonians with local interactions, both gapped and critical \cite{2darea}, and therefore PEPS successfully reproduce the entanglement structure of such quantum states of matter. Fourth, and very differently to what happens for MPS, PEPS can have \emph{infinite correlation length} and hence handle polynomially-decaying correlation functions \cite{dmrgrev1,dmrgrev2,TN}. This means that, at least in theory, one can construct a PEPS with critical properties. Fifth, computing \emph{exact expectation values of an arbitrary PEPS is an exponentially hard problem}, unlike in MPS. One must therefore approximate these calculations, for which different schemes have been developed \cite{TN}. We shall see, however, that recent results in entanglement spectra seem to suggest that PEPS corresponding to gapped $2d$ phases without topological order may be contracted efficiently with large accuracy, yet this is based so far only on particular examples and numerical evidence. Sixth, they have \emph{no canonical form} since the TN in $2d$ has loops, and hence one can not associate left/right orthonormal Schmidt basis simultaneously to every bond index. Still, for non-critical PEPS it is usually possible to find a \emph{quasi-canonical form} \cite{qcform}, which leads to approximate numerical methods for finding ground states (a variation of the so-called ``simple update"  \cite{simple}). Also along this direction, an appropriate choice of gauge in the tensors of a PEPS is known to improve the stability in numerical algorithms \cite{gauge}. 

\subsubsection{TNs with extra dimensions}

In these TNs, the tensors are spanned along the physical spatial dimensions of the underlying lattice, but also along extra ``holographic'' dimensions. Such extra dimensions can be interpreted as a renormalization group direction, which takes into account the description of the physical system at different length scales \footnote{There are TNs with extra dimensions where these are not to be interpreted as coming from a renormalization group. This is the case of, e.g.,  concatenated tensor network states \cite{concat}. However, and for the sake of simplicity, we shall consider here only those TNs with a renormalization origin of their extra dimensions.}. These TNs are thus natural candidates to describe criticality and scale-invariance. 

\emph{\underline{Tree Tensor Networks (TTN).-}}

TTN states are tree-like structures with no loops and one extra dimension, see Fig.\ref{Fig2}(c). The tensors can be understood in terms of coarse-grainings of the physical degrees of freedom at different length scales, where tensors at different layers correspond to different scales. In TTNs tensors may be arbitrary (unconstrained TTNs \cite{unttn}), even though people also use them imposing isometric constrains \cite{isottn}, very much as in the MERA, to be explained later. 

\begin{figure}
\begin{centering}
\includegraphics[width=10cm]{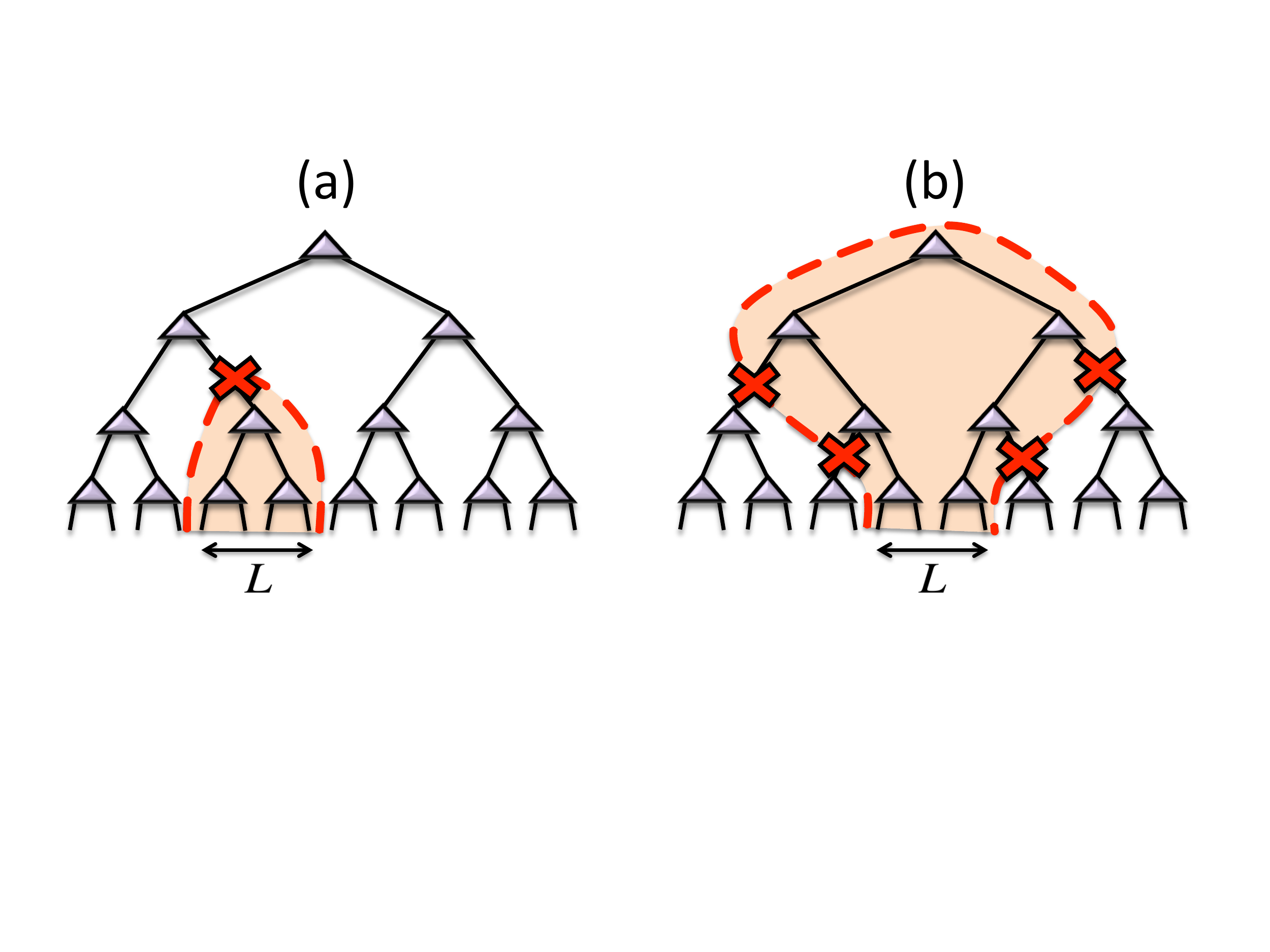} 
\par\end{centering}
\caption{(color online) Entropy of a TTN: (a) partition with $S(L) = O(1)$; (b) partition with $S(L) = O(\log L)$. The entropy is proportional to the number of indices that need to be cut in order to separate the $L$ physical indices of the block from the rest \cite{TN}.} \label{Fig3}
\end{figure}

As for their basic properties, we quickly mention the following. First, \emph{the entanglement entropy of $L$ contiguous spins depends on the bipartition}: some bipartitions obey a $1d$ area-law, whereas some others violate the area-law logarithmically, see Fig.\ref{Fig3}. This means that TTNs, as such, may be a slightly better ansatz for critical $1d$ systems than MPS. People have in fact proposed using linear superpositions of space-translated TTNs -- thus allowing for a logarithmic scaling of the entropy for all blocks -- to simulate $1d$ quantum critical systems \cite{arealaw}.  Second, expectation values can be computed \emph{efficiently} as in an MPS. Third, since there are no loops in the structure, TTNs admit also a \emph{canonical form}, so that every bond index corresponds to orthonormal left/right Schmidt basis (same as in MPS) \cite{ttncan}. Because of their nice analytical properties, TTNs have been used in, e.g., the study of quantum chemistry problems together with MPS approaches \cite{qchem}. 

\emph{\underline{Multi-scale Entanglement Renormalization Ansatz (MERA).-}}

\begin{figure}
\begin{centering}
\includegraphics[width=10cm]{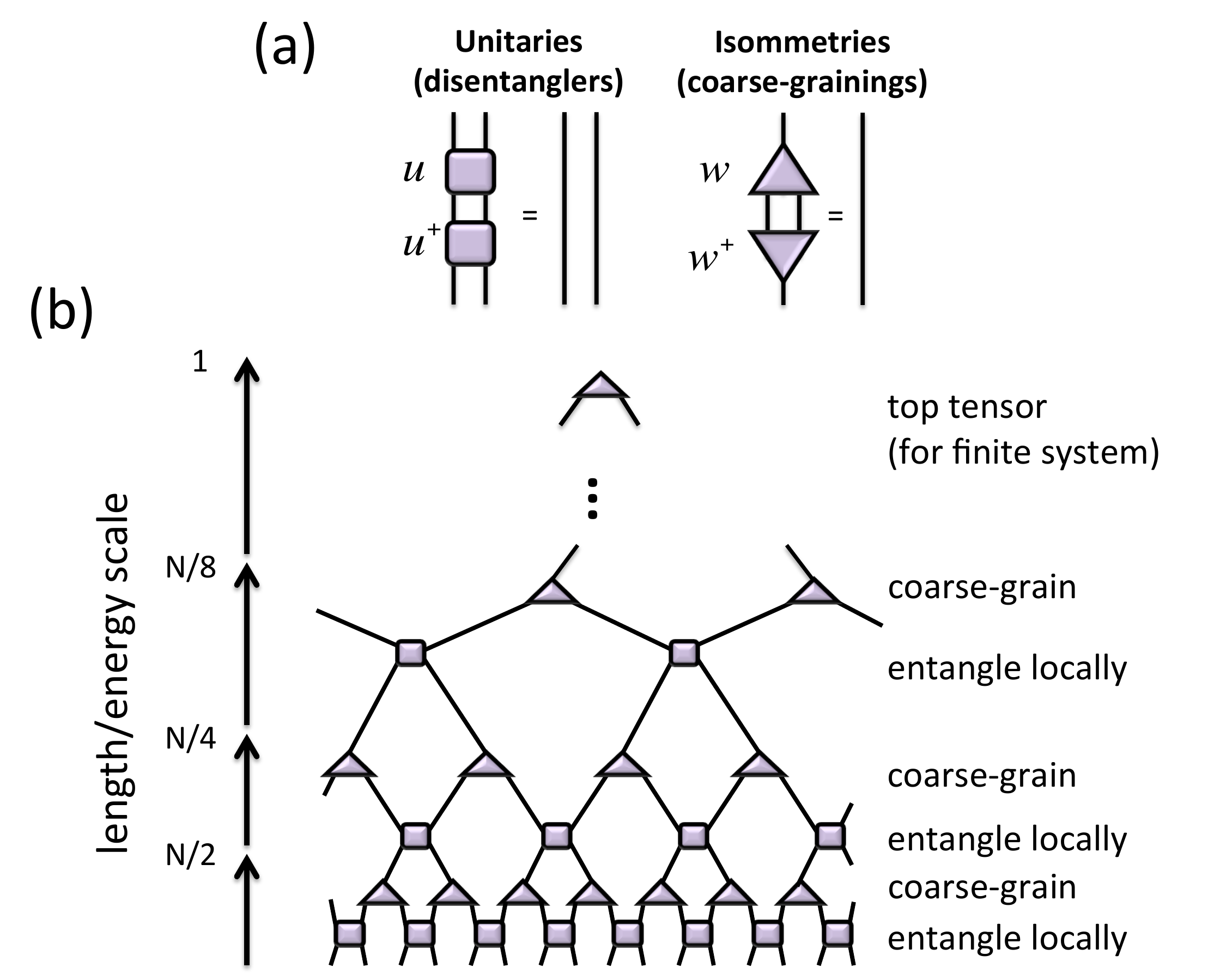} 
\par\end{centering}
\caption{(color online) (a) constraints on MERA tensors; (b) renormalization group interpretation of a $1d$ MERA. This example os for a \emph{binary} MERA, see Refs.\cite{MERA, MERArev}.} \label{Fig4}
\end{figure}

The MERA is a type of TN state where tensors span both along physical directions as well as along a holographic direction, see Fig.\ref{Fig2}(d) for a $1d$ example. This tensor network is at the core of the so-called \emph{Entanglement Renormalization}, and has been widely discussed in the literature. For some detailed introductions to the MERA see, e.g., Refs.\cite{MERA, MERArev}. In a MERA, tensors must obey constraints. Specifically, tensors must be either \emph{unitary} or \emph{isometric}, see Fig.\ref{Fig4}(s). Such constraints allow for a very nice physical interpretation of the MERA: in this tensor network, \emph{entanglement is built locally at all length scales in the system}, see Fig.\ref{Fig4}(b). Unitary tensors entangle neighbouring degrees of freedom at every length scale, whereas isometries coarse-grain the system and preserve the normalisation of the quantum state.

\begin{figure}
\begin{centering}
\includegraphics[width=9cm]{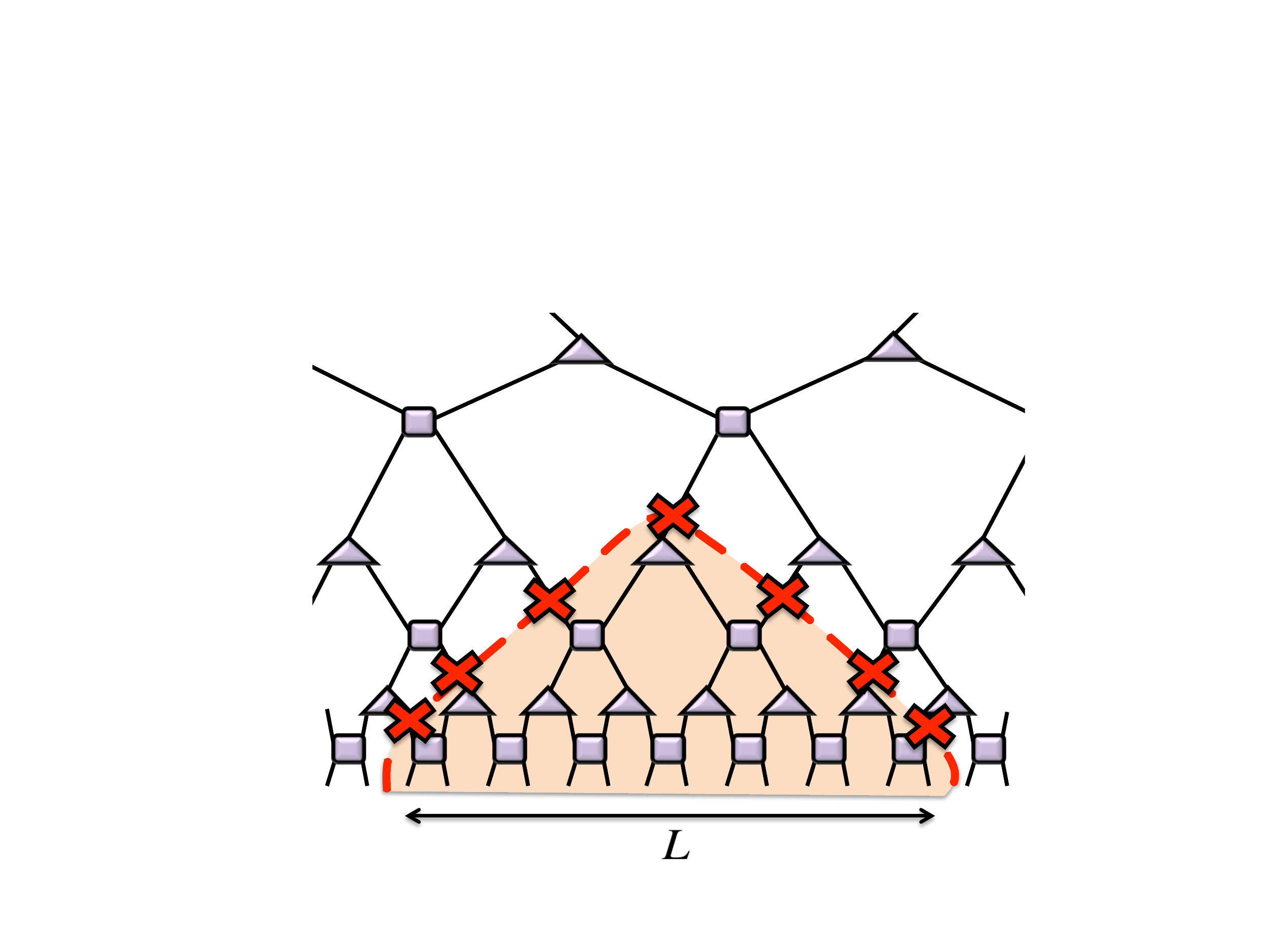} 
\par\end{centering}
\caption{(color online) Entropy of a $1d$ MERA: the number of links to cut in order to disconnect the $L$ physical indices in the block form the rest of the system grows logarithmically with $L$, hence $S(L) = O(\log L)$. This is an example of an area-law in holographic space.} \label{Fig5}
\end{figure}

The MERA is full of nice properties, some of which we quickly remind here. First, the entanglement entropy of a block corresponds to an \emph{area-law in the holographic space}, see Fig.\ref{Fig5}. This means, in particular, that a $1d$ MERA obeys $S(L) = O(\log\chi \log L)$ for a block of $L$ contiguous spins (the exact pre-factors depending on the specifics of the MERA -- binary, ternary... -- \cite{MERA, MERArev}). This is precisely the behaviour predicted by conformal field theory for $1d$ quantum critical systems, therefore the $1d$ MERA correctly captures the structure of entanglement in such quantum states. Second, expectation values and norms can be computed very easily for a MERA because of the constraints in the tensors. For instance, expectation values of local observables at a given site depend only on the tensors inside a \emph{causal cone of bounded width} for the site \cite{MERArev}. Third, as in TTNs, \emph{scale-invariance can be easily implemented} in the MERA just by assuming translation invariance (taking the thermodynamic limit), and choosing the same tensors at every layer (which correspond to different length scales). Finally, numerical algorithms for the MERA must \emph{take into account the constraints} in the tensors \cite{MERArev}. 


\emph{\underline{Branching Multi-scale Entanglement Renormalization Ansatz (branching-MERA).-}}

The branching-MERA was proposed recently \cite{braMERA}Ê as a generalisation of the usual MERA to a different physical scenario. This corresponds to a TN very similar to a usual MERA, but with an additional ingredient: at every renormalization group scale (i.e. at every layer in the holographic direction), one allows for a possible \emph{branching} into independent MERAs, see Fig.\ref{Fig2}(e) for a $1d$ example. Physically this corresponds to the situation where different degrees of freedom get decoupled when flowing along the renormalization group scale. Think, for instance, of spin-charge separation: at short distances spin and charge are coupled, whereas at long distances and in an effective (renormalized) description their Hilbert spaces may decouple into independent ``theories": one for the spin, and one for the charge. But moreover, and regardless of the physical motivation, the branching-MERA is also important because it allows for an arbitrary scaling of the entanglement entropy of a block \cite{braMERA} (even a volume law), yet it maintains all the nice properties of the standard MERA: efficient contraction, causal cone, etc. This, has motivated the branching-MERA as a natural TN ansatz to describe systems that violate the area-law of the entanglement entropy, e.g., $2d$ fermi and spin-bose liquids \cite{fermi,spinbose}. 

In the following sections we will consider different theoretical developments in TN theory with direct implications, both in theory and numerics. For this, it will be useful to keep in mind the quick overview of TN properties that we just did in this section.  

\section{TNs and symmetries: spin networks}

Symmetries are ubiquitous in physics. In fact, they are also one of the fundamental pillars of modern physical theories. What happens, then, if a system to be described by a TN has a symmetry? Are there implications at the level of, say, the entanglement properties and the structure of the TN? Such questions have been discussed in the literature almost since the advent of DMRG. From the perspective of numerical algorithms, it is widely known that the incorporation of symmetries at the TN level can involve huge computational advantages. This has been exploited in DMRG \cite{dmrgsym} and the $1d$ MERA  \cite{merasym0,merasym}Êboth for abelian and non-abelian symmetries, as well as in higher-dimensional systems with abelian symmetries \cite{pepssym}\footnote{To the best of the author's knowledge, non-abelian symmetries are still to be fully exploited in higher-dimensional TN algorithms.}.  Given its importance, here we do a quick review of some of the main implications of symmetries when incorporated in TNs, at a theoretical level. For further details we refer the reader to Refs.\cite{merasym0, weich}. As we shall see, the so-called \emph{spin networks} appear  naturally from TN states with symmetries. Spin networks are used, e.g., in loop quantum gravity to describe quantum states of space  at a certain point in time \cite{spinnetloop}. In this sense, TNs with symmetries turn out to offer a very intriguing first connection between many-body quantum entanglement and gravitation \color{black} via the physics of spin networks \color{black}. 

\subsection{Constraints imposed by a global symmetry}
 
The interested reader is addressed to Ref.\cite{merasym0} (which we follow closely here) for more details. Let us start by assuming that we have symmetry group $G$, compact and completely reducible. There are countless examples of such symmetry groups in physics, say ${\mathbb Z_q}$, $SO(n)$, $U(n)$ and $SU(n)$, to name a few. Let us further assume that $U_g$ corresponds to a unitary matrix representation of an element of the group $g \in G$. Suppose further that we have a quantum state $\ket{\Psi}$ that is \emph{invariant}\footnote{Covariant states, i.e., those that transform according to a specific non-zero charge sector of the symmetry, can also be considered easily.} under a \emph{global} symmetry represented by this group, i.e., under the action of the same group element at every site\footnote{As opposed to a \emph{local} gauge symmetry, where the group element is site-dependent.}. This means
\beq
(U_g)^{\otimes N} \ket{\Psi} = \ket{\Psi} ~~~ \forall g \in G. 
\eeq
The Hilbert space $\mathbb{H}$ at a given physical site decomposes according to a direct sum of the irreducible representations (irreps) of $G$, namely, 
\beq
\mathbb{H} \simeq \bigoplus_a d_a \mathbb{H}^a \simeq \bigoplus_a (\mathbb{D}^a \otimes \mathbb{H}^a) , 
\eeq
with  $\mathbb{H}^a$ the subspace for the irrep with charge $a$, $d_a$ its degeneracy, and $\mathbb{D}^a$ a $d_a$-dimensional degeneracy subspace. Given this decomposition, one can now choose the local basis of the Hilbert space $\mathbb{H}$ at every site as $\ket{a, \alpha_a, m_a}$, with $\alpha_a = 1, 2, \ldots, d_a$ labelling the states in the degeneracy space $\mathbb{D}^a$, and $m_a$ labelling the states in the irrep $\mathbb{H}^a$. By construction, operator $U_g$ also decomposes as a sum of irreps, and in this basis it reads
\beq
U_g = \bigoplus_a (\mathbb{I}^a \otimes U_g^a). 
\label{ug}
\eeq

The above are just fundamental symmetry topics. However, the key result when dealing with TNs is the so-called \emph{Schur's lemma} \cite{schur}. This states that an operator $O$ acting in $\mathbb{H}$ such that $[O, U_g] = 0$ for all $g \in G$ decomposes as
\beq
O = \bigoplus_a (P^a \otimes \tilde{\mathbb{I}}^a), 
\label{schur}
\eeq
i.e., \emph{it can only act non-trivially in the degeneracy subspace}. The proof of Schur's lemma is a basic result in group theory and can be found in many places. Intuitively we can understand this as follows: for $O$ to commute with \emph{all} elements of the group, it must necessarily be the identity in the subspaces of the irreps. However, it can be anything in the degeneracy subspace since, according to Eq.\ref{ug}, $U_g$ is already the identity there. 

The usual approach to deal with symmetries in TNs is to assume that \emph{all tensors are symmetric}. This is, we deal with \emph{a TN made of symmetric tensors}. This ensures, by construction, that the corresponding quantum state will be symmetric. But in any case, it is good to keep in mind that there can also be non-symmetric TN representations of symmetric quantum states which are actually ``cheaper'' (i.e., with smaller bond dimensions) than symmetric ones. For instance, this can happen if the TN has loops, see Ref.\cite{symloops} for specific examples. In the case of $2d$ PEPS, whenever this is injective \cite{injpeps} one can guarantee that a decomposition in terms of symmetric tensors exists. But in any case, it is always a good idea to consider TNs made of symmetric tensors because they may imply several theoretical and computational advantages \cite{merasym0, weich}, as we discuss later. 

\begin{figure}
\begin{centering}
\includegraphics[width=10cm]{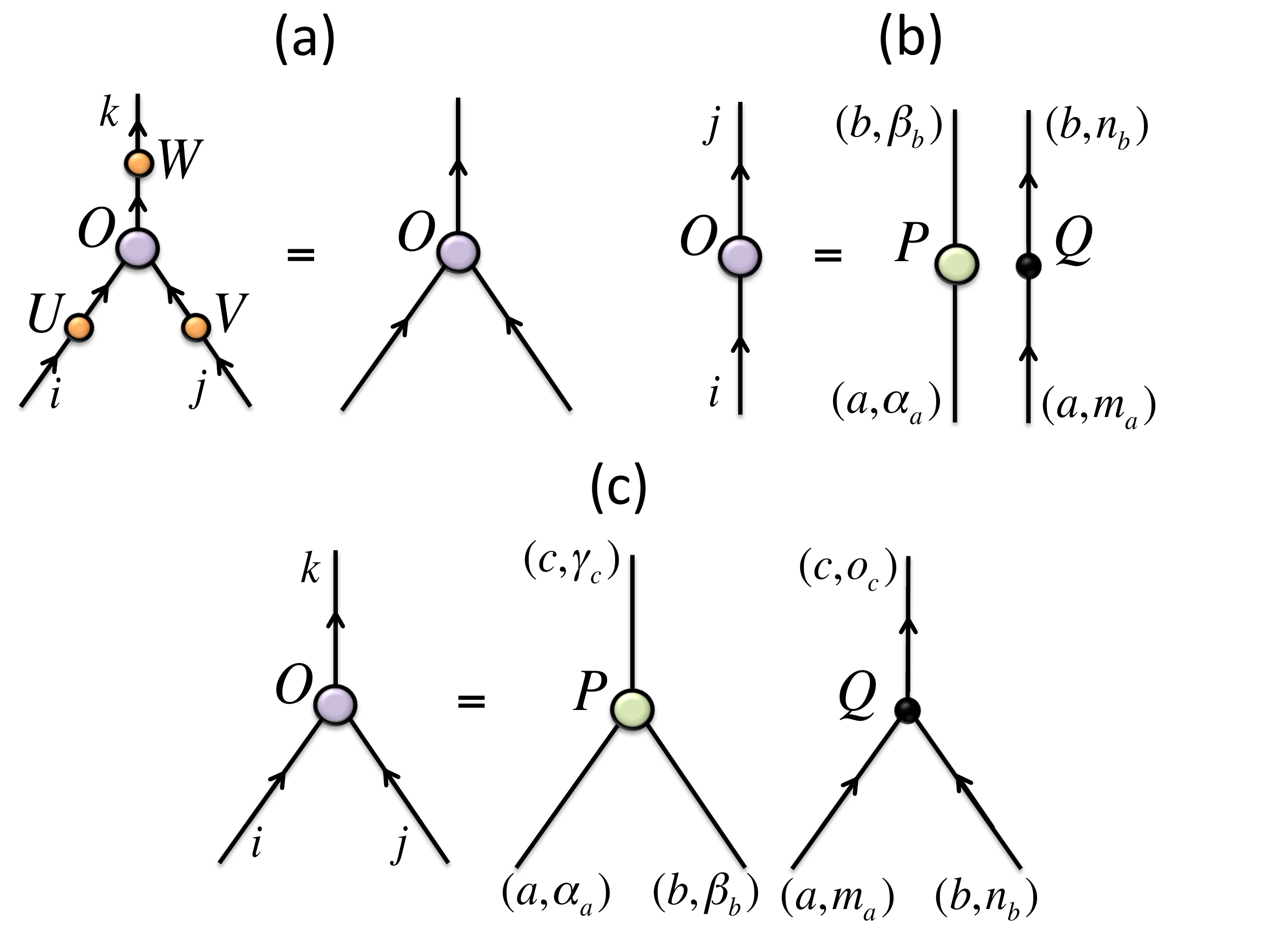} 
\par\end{centering}
\caption{(color online) (a) Symmetric tensor with 3 indices; (b) graphical representation of Eq.\ref{tnsym1}, i.e., Schur's lemma; (c) graphical representation of Eq.\ref{tnsym2}.} \label{Fig7}
\end{figure}

For our purposes, a symmetric tensor with, e.g., 3 indices is one that satisfies the equation represented in the diagram of Fig.\ref{Fig7}(a). Notice that in this diagram we had to use \emph{arrows} in the indices, so that we can specify incoming and outgoing indices whenever the tensor is understood as a matrix after grouping indices. In the figure,  $U$, $V$ and $W$ are unitary matrix representations of a given group element $g$. Notice that these matrix representations may very well be of different sizes, depending on the number of different values that each index can take. \color{black} The expression corresponding to that diagram is 
\beq
(U\otimes V) O W = O, 
\label{uvow}
\eeq
where incoming indices in the diagram are understood as left-hand-side matrix indices in the equation. \color{black}

The important thing to realise is that, using ultimately Schur's lemma, one can arrive to the result that symmetric tensors can be decomposed in two pieces: one completely determined by the symmetry acting on the irrep subspaces (similar to the identity in Eq.\ref{ug}), and another which contains the degrees of freedom acting entirely on the degeneracy subspace. For a tensor with two legs, i.e., a matrix, this is readily seen by Schur's lemma in Eq.\ref{schur}. In components, labeling $i \equiv (a, \alpha_a, m_a)$ and $j \equiv (b, \beta_b, n_b)$, one has that 
\beq
O_{ij} = (P^{ab})_{\alpha_a \beta_b} (Q^{ab})_{m_a n_b}, 
\label{tnsym1}
\eeq
with $Q^{ab} = \delta_{ab} \delta_{m_a n_b}$. This is represented in the diagram of Fig.\ref{Fig7}(b). Let us interpret this equation: it means that for fixed values of the irrep charges $a$ and $b$, the matrix $O_{ij}$ splits into a degeneracy tensor $P^{ab}$, where only the piece with $a=b$ contributes, and another tensor $Q^{ab}$. The degeneracy tensor contains all the degrees of freedom of $O_{ij}$, whereas $Q^{ab}$ is completely determined by the symmetry. 

One can generalise this result further. For instance, for the case of a tensor with three legs, one needs to take into account that the tensor product of two irreps can be decomposed as a sum of irreps. Say, for the product of two irreps with charges $a$ and $b$ we have that $\mathbb{H}^a \otimes \mathbb{H}^b \simeq \bigoplus_c N^c_{ab} \mathbb{H}^c$, with $N^c_{ab}$ the number copies of irrep $c$ appearing in the decomposition. Groups such as SU(3) have multiplicities, i.e., $N^c_{ab} > 1$ for some $a,b$ and $c$. However, most treatments of symmetries in TN states assume, for the sake of simplicity, that the group $G$ is multiplicity-free, i.e., $N^c_{ab} \le 1$. While not being completely general, this case already includes important examples such as $U(1)$ and $SU(2)$. Here we will make also this assumption, for the sake of simplicity\footnote{Groups with multiplicity can also be considered easily \cite{merasym0, weich}.}. Let us call $(I^{abc})_{m_a n_b o_c}$ the change of basis between the product basis $\ket{a, m_a} \otimes \ket{b, n_b}$ and the coupled basis $\ket{c, o_c}$, i.e., the Clebsch-Gordan coefficients of the group. The well-known Wigner-Eckart theorem implies that a symmetric tensor $O$ with, say, two incoming indices $i,j$ and one outgoing index $k$ can be decomposed as 
\beq
O_{ijk} = (P^{abc})_{\alpha_a \beta_b \gamma_c} (Q^{abc})_{m_a n_b o_c}, 
\label{tnsym2}
\eeq
where again $P^{abc}$ is a degeneracy tensor that contains all the degrees of freedom of $O_{ijk}$, and the structural tensor $Q^{abc}$ is completely fixed by the symmetry group (Clebsch-Gordan coefficients). This is represented in the TN of Fig.\ref{Fig7}(c). 

From here on we could continue decomposing tensors with more and more indices, i.e., 4 indices, 5 indices, and so on. Even if this is possible \cite{merasym0}, in the end this is not entirely necessary because all TN states can always be written in terms of tensors with three indices at most\footnote{This can be achieved using, e.g., appropriate tensor decompositions such as the singular value and $QR$ decompositions.}. Thus, Eq.\ref{tnsym1} and Eq.\ref{tnsym2} are the fundamental ones to be used when dealing with complex structures in TN states with symmetries.  

\begin{figure}
\begin{centering}
\includegraphics[width=10cm]{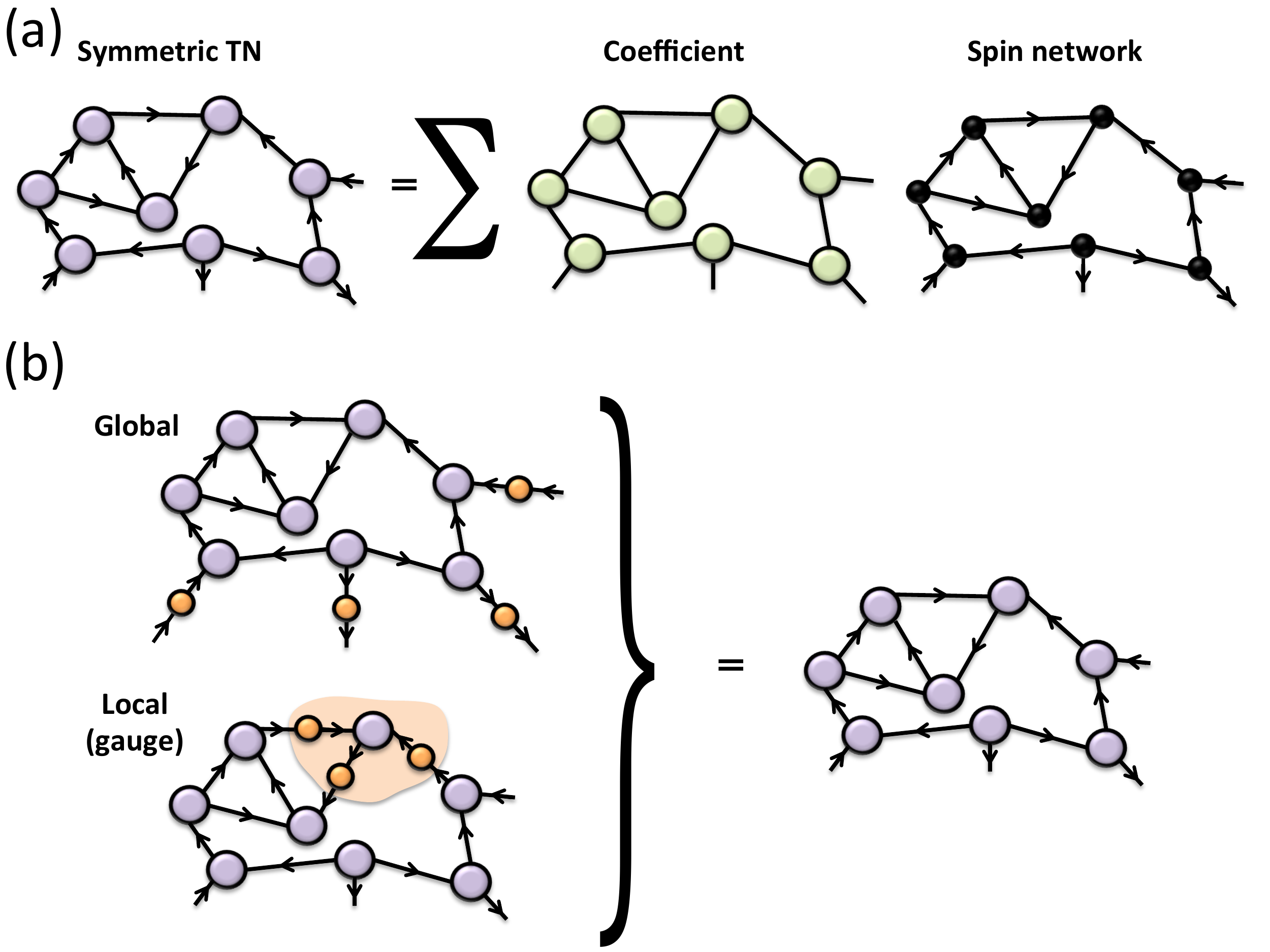} 
\par\end{centering}
\caption{(color online) (a) A symmetric TN decomposes as the sum of exponentially-many spin networks, where the coefficients are given by TNs of degeneracy tensors. In the diagram, the sum runs over all contracted indices (with the same value of charges for the coefficient network and spin network at every term in the sum); (b) using a construction where all the tensors are symmetric, the resultant TN has a global symmetry on the open (physical) indices -- where the same element of the group acts on all physical sites --, but also a local gauge symmetry in the individual tensors -- where different group elements could act locally on individual tensors --.}
\label{Fig8}
\end{figure}

What happens now to the whole TN state? If each tensor in a TN decomposes in a $(P,Q)$ form as in the previous equations then the net result is that, for a fixed value of the charges labelling the irreps, the TN factorizes itself in two pieces: a TN of degeneracy tensors, and a TN of structural tensors. The second piece is a directed graph (i.e., with arrows), whose vertices correspond to intertwining operators of $G$ such as Clebsch-Gordan coefficients, and whose edges are labelled by irreps. And this is nothing but a \emph{spin network}, which is a well-known object in mathematical physics and loop quantum gravity, where it is used to describe quantum states of space at a given time. Thus, a symmetric TN for a quantum state $\ket{\Psi}$ of $N$ sites can be understood as a superposition of exponentially many spin networks with $N$ open indices. The coefficients of the superposition are given by TNs of degeneracy tensors, see Fig.\ref{Fig8}(a). 

\subsection{Advantages of the symmetric description}

The description of a symmetric quantum state in terms of a TN of symmetric tensors is clearly very appealing both from the theoretical and numerical points of view. The reasons for this are: 

\begin{enumerate} 

\item{\emph{It allows to simulate systems with specific quantum numbers.} For instance, for $SU(2)$-symmetric systems, one can target quantum states with total spin $j=0$ for a singlet state, or total spin $j=1$ for, e.g., a triplet excitation. For $U(1)$-symmetric systems, one can target states with a definite component of the angular momentum in the $z$ direction or, similarly, states with a well-defined number of particles.}

\item{\emph{It allows for a more compact description of the TN}, where all the degrees of freedom are in the tensors $P^{abc}$. In fact, notice that a large amount of the entanglement in the system is directly given by tensors $Q^{abc}$, and hence by the symmetry constraints only.}

\item{\emph{It allows for lower computational costs in numerical algorithms.} Here the advantage comes from two fronts: first, calculations such as matrix multiplication can be done symmetry sector-by-symmetry sector (i.e. block-by-block), which brings a computational advantage. Matrix decompositions such as singular value decompositions can also be done more efficiently, while  reshaping (or grouping) of indices in tensors usually adds a small sub-leading extra cost that does not typically overcome the overall computational speed-up \cite{merasym0, merasym}. Second, many of the manipulations in TN methods are based on repeated iterations of some basic step (examples of this are, e.g., the iPEPS \cite{iPEPS} and iTEBD \cite{itebd} algorithms). This, in turn, involve many repetitions of the same manipulations of structural tensors. Such manipulations can be computed only once, and their result can be stored in memory and recycled for later use during the rest of the algorithm. While this clearly produces an increase in memory usage, it also reduces the number of operations. In practice, the level up to which such a ``pre-computation'' is really worthy depends on the specifics of (i) the method to be programmed, and (ii) the computer platform to be used.}

\item{Finally, an advantage from the theoretical perspective is that \emph{TNs of symmetric tensors are also, by construction, gauge invariant.} For instance see the TN from Fig.\ref{Fig8}(b). If such a TN is to be interpreted as a lattice version of some space geometry (as we shall argue for MERA in Sec.6), this will automatically inherit a gauge symmetry in the bulk, see Ref.\cite{merasym}. \color{black}ÊNotice, however, that this local gauge invariance refers only to the virtual degrees of freedom, and not to the local gauge invariance in the usual sense, i.e., where independent gauge transformations can be applied to physical sites \cite{gaug}. \color{black}}

\end{enumerate}

\subsection{A simple example}

Let us illustrate the drastic reduction of parameters with a simple example. Consider a 3-index tensor with bond dimensions 3, 3 and 9. If the tensor is non-symmetric, then the number of complex parameters in the tensor is $3 \times 3 \times 9 = 81$. However, if the tensor is $SU(2)$-invariant, the number of degrees of freedom drops down by a large factor. To see this, imagine that the first two indices carry each one an irrep with total spin 1 each, with degeneracy 1, and 3 possible states within each spin-1 irrep. Composing irreps of angular momentum we know that $1 \otimes 1 = 0 \oplus 1 \oplus 2$. Therefore, the third index can carry irreps 0 with dimension 1, 1 with dimension 3, and 2 with dimension 5, all without degeneracies. Now, having a look at Eq.\ref{tnsym2}, one sees that tensor $O^{abc}$ has no degeneracy indices in our case and only three non-zero components, say, $O^{110}$, $O^{111}$ and $O^{112}$. Thus, the $SU(2)$-symmetric version of the 3-index tensor is determined entirely by 3 complex parameters only. Compared to the original 81 complex parameters of a non-symmetric tensors, this is a reduction of more than 90\% of the degrees of freedom! 

\color{black}
To conclude this section let us mention that, in spite of the obvious power of the symmetric description of tensors (as exemplified above), there may still exist the problem of identifying the set of irreps that need to be included in each of the tensor indices in order to attain a sufficiently accurate description of a quantum many-body state. This factor may affect the efficiency of related tensor network methods, and needs to be considered when developing specific numerical algorithms. 
\color{black}

\subsection{Recent implementations}

Symmetries have been widely used in DMRG and other MPS techniques to lower the computational costs of the algorithm, so that very large bond dimensions can be achieved. There are countless examples of this, both for abelian and non-abelian symmetries, see e.g., Refs.\cite{dmrgsym}. For the $1d$ MERA, abelian and non-abelian symmetries have also been considered \cite{merasym}. In Ref.\cite{symloops} and analysis of the symmetric bond dimension versus the minimal bond dimension of a TN was also done in the context of the $1d$ MERA. A general framework for symmetries in TNs is also provided in Ref.\cite{weich}. For $2d$ PEPS abelian symmetries have been implemented at the algorithmic level \cite{pepssym}, whereas non-abelian symmetries remain as an open computational challenge. Moreover, in the context of the simulation fermionic systems with TNs, ${\mathbb Z}_2$ parity symmetry has also been used in, e.g. $2d$ MERA \cite{fmera} and $2d$ PEPS \cite{fipeps} algorithms. From a theoretical perspective, relation between symmetries in TNs, string/brane order, and topological order, has also been nicely addressed in a number of  contributions \cite{injpeps, david}. 

\section{TNs and fermions: graphical projections}

The accurate simulation of fermionic systems in $2d$ is one of the most important open problems in condensed matter physics. For instance, the phase diagram of the fermionic Hubbard and $t-J$ models \cite{hub}, believed to be relevant for high-temperature superconductivity, are still unsolved to a large extent in spite of many years of study. Not only that: according to the standard model of particle physics, everyday matter is made of fundamental fermions, say, leptons and quarks. It looks then natural that simulation methods should be able to tackle these systems. Still, fermionic Hamiltonians remain as some of the hardest models to simulate in higher dimensions, mainly because of the so-called sign problem in quantum Monte Carlo \cite{sign}. While other approaches do not have such a sign problem, they are also limited by other factors. 

TN methods can be adapted to simulate fermionic systems in any dimension, thus offering a promising numerical alternative. Several approaches have been developed in order to implement the fermionic statistics at the level of TN algorithms \cite{fmera, fipeps, fPEPS, pit, foc}, in the end being all equivalent. Here we quickly review the approach taken in Ref.\cite{fipeps}, which can be very nicely stated entirely in terms of graphical rules and tensor diagrams. We omit here many formal derivations, so the interested reader is addressed to Ref.\cite{fipeps} for more information. \color{black} Moreover, we also briefly sketch the equivalence of this approach to other methods to deal with fermions in tensor network states. \color{black} 

From a logical point of view, one should start from the second quantisation formalism for fermions, and then see how this translates into the TN language. Of course this is what has been done historically in the development of fermionic TN algorithms. However, for the sake of this paper we take a different approach: we first start from the actual solution to the problem, and then we see via specific examples why this is correct. 

\subsection{TN fermionisation rules} 

\begin{figure}
\begin{centering}
\includegraphics[width=12cm]{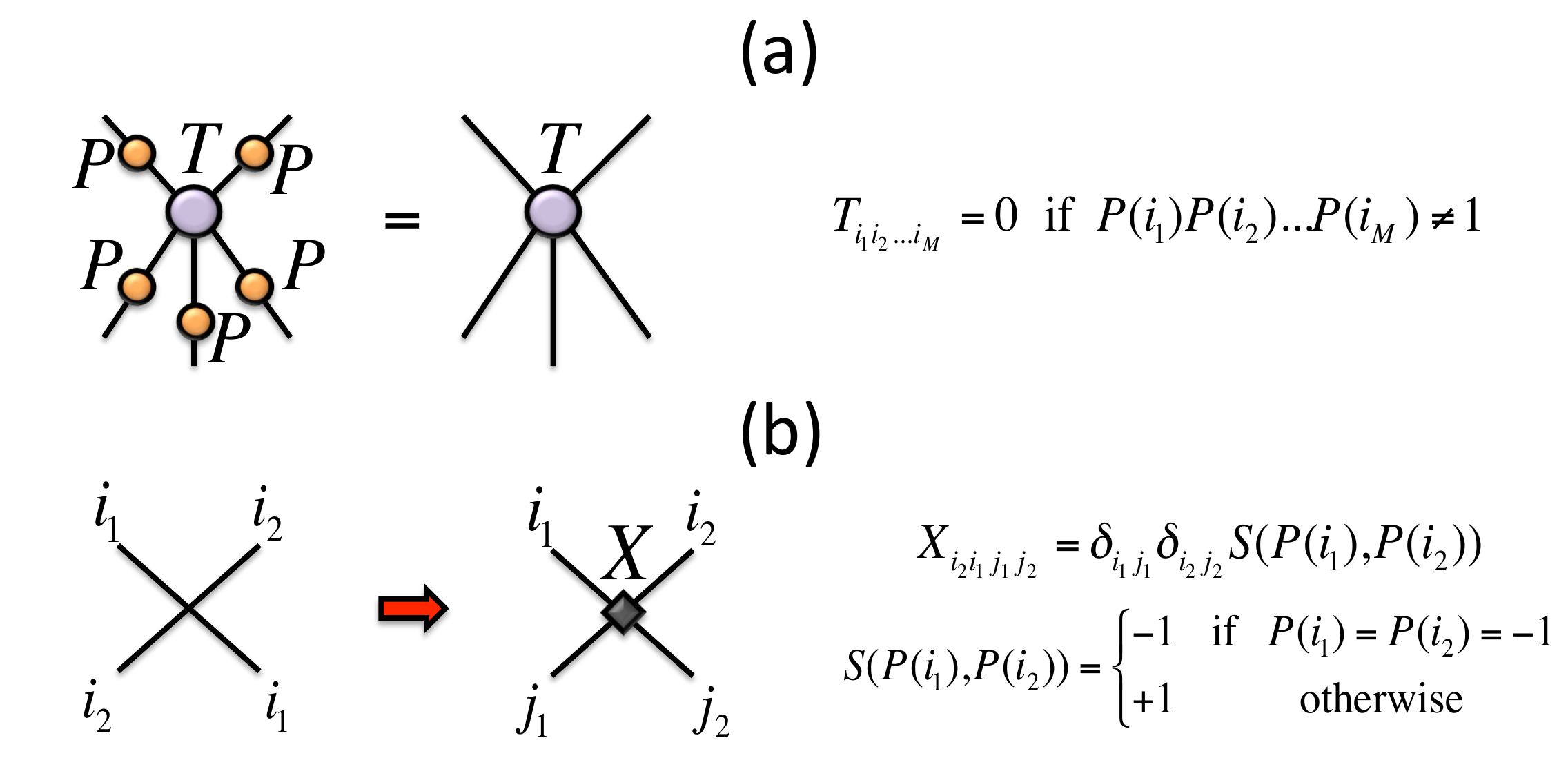} 
\par\end{centering}
\caption{(color online) TN fermionization rules: (a) parity-invariance of tensors, where $P$ is a representation of the $\mathbb{Z}_2$ parity symmetry operator; (b) crossings are replaced by fermionic swaps.}
\label{Fig10}
\end{figure}

The ``fermionisation'' of a TN method is based on the following two rules, according to the approach in Ref.\cite{fipeps}: 

\begin{enumerate} 
\item{\emph{Use parity-symmetric tensors}, see Fig.\ref{Fig10}(a). Fermionic parity, i.e., whether the total number of fermions is even or odd, is a good $\mathbb{Z}_2$ symmetry for fermionic systems. Thus, it is always a good idea to incorporate it directly at the level of the TN (see previous section).  We will see that parity symmetry in the tensors allows to do some important manipulations in the TN.}

\item{\emph{Replace line crossings in the planar representation of the TN by fermionic swap gates}, which are defined as in Fig.\ref{Fig10}(b). It is at this point where the fermionic statistics comes into play. This replacement has the following physical interpretation: every wire or line in the TN diagram represents a fermionic degree of freedom, either ``physical'' (such as the physical indices) or ``virtual'' (such as the bond indices). In practice, when an odd number of fermions (odd parity) swap their oder with an odd number of fermions, the wave-function gets multiplied by $-1$. This is what the fermionic swap takes into account, by reading the parity ``charge'' of each index and multiplying by $-1$ whenever appropriate.} 
\end{enumerate}

At this level it may not be obvious at all why these two rules are correct. To justify them, we now discuss a number of examples of increasing complexity that will in fact illustrate that these two rules, when combined together, turn out to give the correct TN description of a fermionic system in a completely graphical and diagrammatic way.

\subsection{Three examples}

Let us start by reminding that in second quantisation one must choose an ordering for the fermions, i.e., an ordering on how fermionic operators are applied to a fermionic vacuum. Whereas for $1d$ systems there is a natural ordering, this is not the case for higher-dimensional systems. In any case, fermionic ordering is \emph{not} a physical property since it is chosen by us. Therefore, physical quantities such as expectation values in the end do not depend on it. Here we shall see how different choices in the fermionic ordering lead to different graphical projections of the TN on a plane, when translated into the language of TN diagrams. The relative minus signs coming from permuting the fermions from one order to another are taken into account by the relative difference in fermionic swap gates between the two corresponding graphical projections. 

To define properly the operators, we take into account the parity ${\mathbb Z}_2$ symmetry. Following the previous section, we can see that because of the symmetry, states in the different local Hilbert spaces can be labeled as $\ket{i} \equiv \ket{a, \alpha_a}$, with $a$ the irrep of the parity symmetry ($a = 1$ for even, $a = -1$ for odd), and $\alpha_a$ a degeneracy index (all irreps are one-dimensional). Hence, one can identify two types of operators: those that preserve the parity of a given state, and those that change it. Let us call these operators as follows: 
\beqa
a^{\dagger}_{(1,\alpha_1)} &~~~~& {\rm parity ~ preserving} \nonumber \\ 
a^{\dagger}_{(-1,\alpha_{-1})} &~~~~& {\rm parity ~changing}. 
\eeqa 
For instance, for polarised spinless fermions we would have $a^{\dagger}_{(1,1)}  = {\mathbb I}$ and $a^{\dagger}_{(-1,1)} = b^{\dagger}$, with $b^{\dagger}$ the usual fermionic creation operator. But this notation can handle more complicated situations, e.g.,  when we have fermions with internal degrees of freedom such as spin or colour. For instance, for fermions with spin-$1/2$ (as in the Hubbard model) one has
\beqa
a^{\dagger}_{(1,1)} = {\mathbb I} &~~~~~& a^{\dagger}_{(1,2)} = b^{\dagger}_{\uparrow}  b^{\dagger}_{\downarrow} \nonumber \\
a^{\dagger}_{(-1,1)} = b^{\dagger}_{\uparrow}  &~~~~~& a^{\dagger}_{(-1,2)} = b^{\dagger}_{\downarrow}, 
\eeqa 
with $b^{\dagger}_{\uparrow / \downarrow}$ the operator that creates a fermion with spin up / down respectively. Let us now write these operators as $a^{\dagger}_{i}$, where $i$ is to be understood as the mode to be created. With this notation, the operators satisfy the commutation relations 
\beq
a^{\dagger}_{i}a^{\dagger}_{j} = S(i,j) a^{\dagger}_{j}a^{\dagger}_{i} ~~~~~~a^{\dagger}_{i}a_{j} = S(i,j) a_{j}a^{\dagger}_{i}, 
\eeq
with $S(i,j)$ the fermionic swap operator in Fig.\ref{Fig10}(b). The language in terms of these operators is convenient, since it allows us to correctly implement modes with complete generality, including physical and virtual degrees of freedom. Sometimes we call these \emph{generalised parity operators}. 

After having fixed the notation, we consider here three simple examples of increasing complexity that will allow us to see how the above fermionisation rules of a TN are correct and why they can be generalised to arbitrary systems. In increasing level of complexity, we consider below the case of a 2-site MPS for 2 fermions, 3-site MPS for 3 fermions with periodic boundary conditions, and 4-site MPS for 4 fermions with periodic boundary conditions. 

\begin{enumerate}
\item{\emph{2-site MPS for 2 fermions.}  We start by assigning a normal ordering to the fermions. Now, we consider the parity-preserving fermionic operators 
\beq
 T^{[1]}_{i_1 \mu} a^\dagger_{i_1}a_{\mu}, ~~~~~ T^{[2]}_{\mu i_2}a^\dagger_{\mu}a_{i_2}^\dagger
\eeq
where $a_x$ and $a^\dagger_x$ are creation and annihilation operators for mode $x$ as defined above. The fermionic degrees of freedom $i_1$ and $i_2$ correspond to the actual physical fermions, whereas $\mu$ corresponds to a virtual mode. A graphical representation of these two operators is in Fig.\ref{Fig11}. Next, we can just take the matrix product of both operators giving
\beq
\sum_{\mu} T^{[1]}_{i_1 \mu} T^{[2]}_{\mu i_2} a^\dagger_{i_1}a_{\mu}a^\dagger_{\mu}a_{i_2}^\dagger
 =  \sum_{\mu} T^{[1]}_{i_1 \mu} T^{[2]}_{\mu i_2} a^\dagger_{i_1}a_{i_2}^\dagger. 
 \eeq
When acting on a fermionic vacuum $\ket{\Omega}$, one actually gets a 2-site MPS quantum state: 
\beq
\ket{\Psi} = \sum_{i_1  i_2} \left( \sum_\mu T^{[1]}_{i_1 \mu} T^{[2]}_{\mu i_2}\right) a^\dagger_{i_1}a_{i_2}^\dagger  \ket{\Omega}. 
\label{2mp}
\eeq
This is represented in Fig.\ref{Fig11}(a). Thus, the bond dimension of the MPS corresponds to a generalised parity degree of freedom that is traced out in order to get the final quantum state. Now, if we swap the position of the two physical fermions in real space, the overall wave-funcion is given by 
\beq
SWAP_{12} \ket{\Psi} =  \sum_{i_1  i_2} \left( \sum_\mu T^{[1]}_{i_1 \mu} T^{[2]}_{\mu i_2} S(i_1,i_2)\right) a_{i_2}^{\dagger}a^\dagger_{i_1}  \ket{\Omega}, 
\eeq
where $SWAP_{12}$ is an an operator that swaps the physical fermions $1$ and $2$. Notice that as a consequence of the swap and the fermionic anticommutation rules, the fermionic swap tensor $S(i_1,i_2)$ shows up. We can represent this diagrammatically as in Fig.\ref{Fig11}(b) where we represent a swap of the fermions by crossing their corresponding lines in the diagram, \emph{and} add $S(i_1,i_2)$ by introducing the fermionic swap operator from Fig.\ref{Fig10}(b) right at the crossing point. Thus, a change in the fermionic ordering of physical fermions can be accounted for by fermionic swap operators in TN diagrams.}
\begin{figure}
\begin{centering}
\includegraphics[width=12cm]{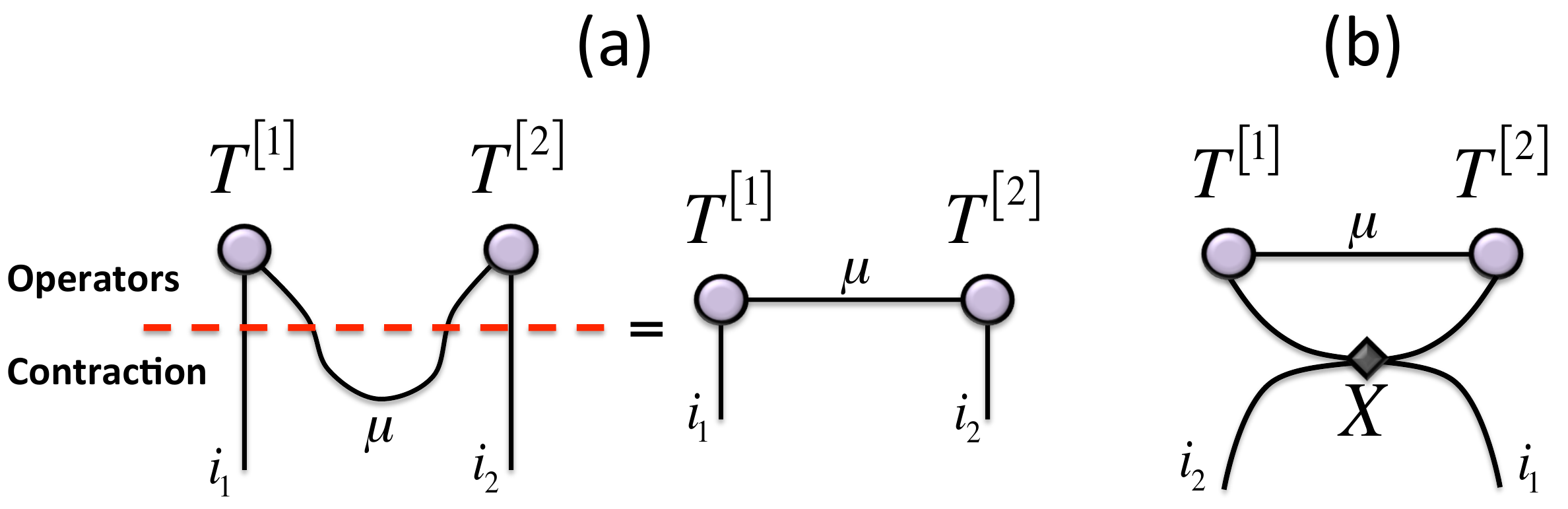} 
\par\end{centering}
\caption{(color online) Fermionic TN diagram for a 2-site MPS.}
\label{Fig11}
\end{figure}

\item{\emph{3-site MPS for 3 fermions with periodic boundary conditions}. This is the case from Fig.\ref{Fig12}. As before, we start by considering three parity-preserving operators 
\beq
T^{[1]}_{i_1 \mu \nu} a^\dagger_{i_1}a_{\mu} a_{\nu} ~~~~~T^{[2]}_{\nu i_2 \rho}a^\dagger_{\nu}a^{\dagger}_{i_2}a_{\rho} ~~~~~ T^{[3]}_{\rho \mu i_3}a^\dagger_{\rho} a^\dagger_{\mu}a^{\dagger}_{i_3}, 
\eeq
where fermionic modes are sorted according to a normal ordering. Taking a trace of the virtual degrees of freedom and applying the result to the vacuum, we get the quantum state 
\beqa
\ket{\Psi} &=& \sum_{i_1 i_2 i_3} \left(\sum_{\mu \nu \rho} T^{[1]}_{i_1 \mu \nu} T^{[2]}_{\nu i_2 \rho} T^{[3]}_{\rho \mu i_3} \right) a^\dagger_{i_1}a_{\mu} a_{\nu}  a^\dagger_{\nu}a^{\dagger}_{i_2}a_{\rho} a^\dagger_{\rho} a^\dagger_{\mu}a^{\dagger}_{i_3} \ket{\Omega} \nonumber \\
&=& \sum_{i_1 i_2 i_3} \left(\sum_{\mu \nu \rho} T^{[1]}_{i_1 \mu \nu} T^{[2]}_{\nu i_2 \rho} T^{[3]}_{\rho \mu i_3} \right) a^\dagger_{i_1}a_{\mu} a^{\dagger}_{i_2} a^\dagger_{\mu}a^{\dagger}_{i_3} \ket{\Omega} \nonumber \\
&=& \sum_{i_1 i_2 i_3} \left(\sum_{\mu \nu \rho} T^{[1]}_{i_1 \mu \nu} T^{[2]}_{\nu i_2 \rho} T^{[3]}_{\rho \mu i_3} S(i_2,\mu) \right) a^\dagger_{i_1}a^{\dagger}_{i_2} a^{\dagger}_{i_3} \ket{\Omega} ,
\eeqa
where in the last line we anticommuted operators $a_{\mu}$ and $a^{\dagger}_{i_2}$ and therefore we got the fermionic swap $S(i_2,\mu)$.  This state corresponds to the diagram in Fig.\ref{Fig12}, where we can see in a graphical way the appearance of the fermionic swap operator that swaps modes $i_2$ and $\mu$. The resulting state, as seen in the figure, is nothing but an MPS with periodic boundary conditions.} 
\begin{figure}
\begin{centering}
\includegraphics[width=12cm]{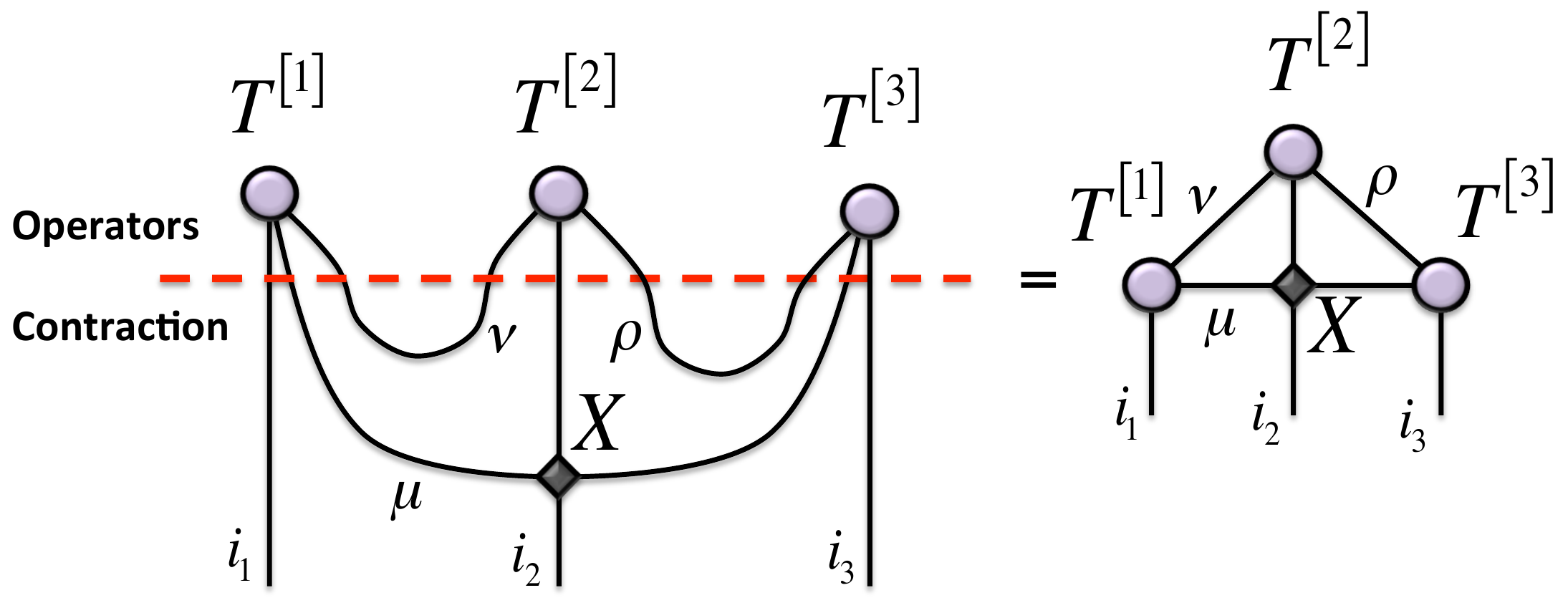} 
\par\end{centering}
\caption{(color online) Fermionic TN diagram for a 3-site MPS with periodic boundary conditions.}
\label{Fig12}
\end{figure}

\item{\emph{4-site MPS for 4 fermions with periodic boundary conditions.} This corresponds to the case depicted in Fig.\ref{Fig13}. We start from four parity-preserving operators
\beq
T^{[1]}_{i_1 \mu \omega} a^\dagger_{i_1}a_{\mu} a_{\omega} ~~~~~T^{[2]}_{\mu i_2 \nu}a^\dagger_{\mu}a^{\dagger}_{i_2}a_{\nu} ~~~~~ T^{[3]}_{\rho \nu i_3}a^\dagger_{\rho} a^\dagger_{\nu}a^{\dagger}_{i_3}  ~~~~~T^{[4]}_{\omega i_4 \rho}a^\dagger_{\omega}a^{\dagger}_{i_4}a_{\rho}
\eeq
where again we have chosen a normal ordering for the different fermionic modes. We now trace out the virtual degrees of freedom as in Fig.\ref{Fig13}. Notice that this time, the ordering that we choose for the four physical fermions does not coincide with the labelling of the tensors. In this case we get the quantum state  
\beqa
\ket{\Psi} &=& \sum_{i_1 i_4 i_2 i_3} \left(\sum_{\mu \nu \rho \omega} T^{[1]}_{i_1 \mu \omega} T^{[4]}_{\omega i_4 \rho} T^{[2]}_{\mu i_2 \nu} T^{[3]}_{\rho \nu i_3} \right) a^\dagger_{i_1}a_{\mu} a_{\omega} a^\dagger_{\omega}a^{\dagger}_{i_4}a_{\rho} a^\dagger_{\mu}a^{\dagger}_{i_2}a_{\nu} a^\dagger_{\rho} a^\dagger_{\nu}a^{\dagger}_{i_3} \ket{\Omega}  \nonumber \\ 
& = & \sum_{i_1 i_4 i_2 i_3} \left(\sum_{\mu \nu \rho \omega} T^{[1]}_{i_1 \mu \omega} T^{[4]}_{\omega i_4 \rho} T^{[2]}_{\mu i_2 \nu} T^{[3]}_{\rho \nu i_3} \right) a^\dagger_{i_1}a_{\mu} a^{\dagger}_{i_4}a_{\rho} a^\dagger_{\mu}a^{\dagger}_{i_2}a_{\nu} a^\dagger_{\rho} a^\dagger_{\nu}a^{\dagger}_{i_3} \ket{\Omega} \nonumber \\ 
& = & \sum_{i_1 i_4 i_2 i_3} \left(\sum_{\mu \nu \rho \omega} T^{[1]}_{i_1 \mu \omega} T^{[4]}_{\omega i_4 \rho} T^{[2]}_{\mu i_2 \nu} T^{[3]}_{\rho \nu i_3} S(\mu,i_4) S(\rho,\mu) S(\rho,i_2) S(\rho,\nu) \right) a^\dagger_{i_1}a^{\dagger}_{i_4}a^{\dagger}_{i_2}a^{\dagger}_{i_3} \ket{\Omega} \nonumber \\ 
& = & \sum_{i_1 i_4 i_2 i_3} \left(\sum_{\mu \nu \rho \omega} T^{[1]}_{i_1 \mu \omega} T^{[4]}_{\omega i_4 \rho} T^{[2]}_{\mu i_2 \nu} T^{[3]}_{\rho \nu i_3} S(\mu,i_4)  \right) a^\dagger_{i_1}a^{\dagger}_{i_4}a^{\dagger}_{i_2}a^{\dagger}_{i_3} \ket{\Omega} .
\eeqa

Let us understand the above equation. Until the third line we simply used the fermionic anti-commutation relations. However, in order  to jump from the third to the fourth line, we used the parity symmetry of operator $T^{[2]}$. Indeed, because of the parity symmetry of this operator it is quite easy to see that $S(\rho,\mu) S(\rho,i_2) S(\rho,\nu) =1$ always, so that we get the final expression in the last line. This is shown in the last diagram of Fig.\ref{Fig13}, where we see that the parity symmetry of $T^{[2]}$ allows us to make a so-called \emph{jump move}, moving the wire for index $\rho$ over tensor $T^{[2]}$ and getting the final diagram. This diagram is nothing but a fermionic MPS with periodic boundary conditions, and corresponds exactly to the quantum state in the last line of the previous equation.}
\end{enumerate}
\begin{figure}
\begin{centering}
\includegraphics[width=12cm]{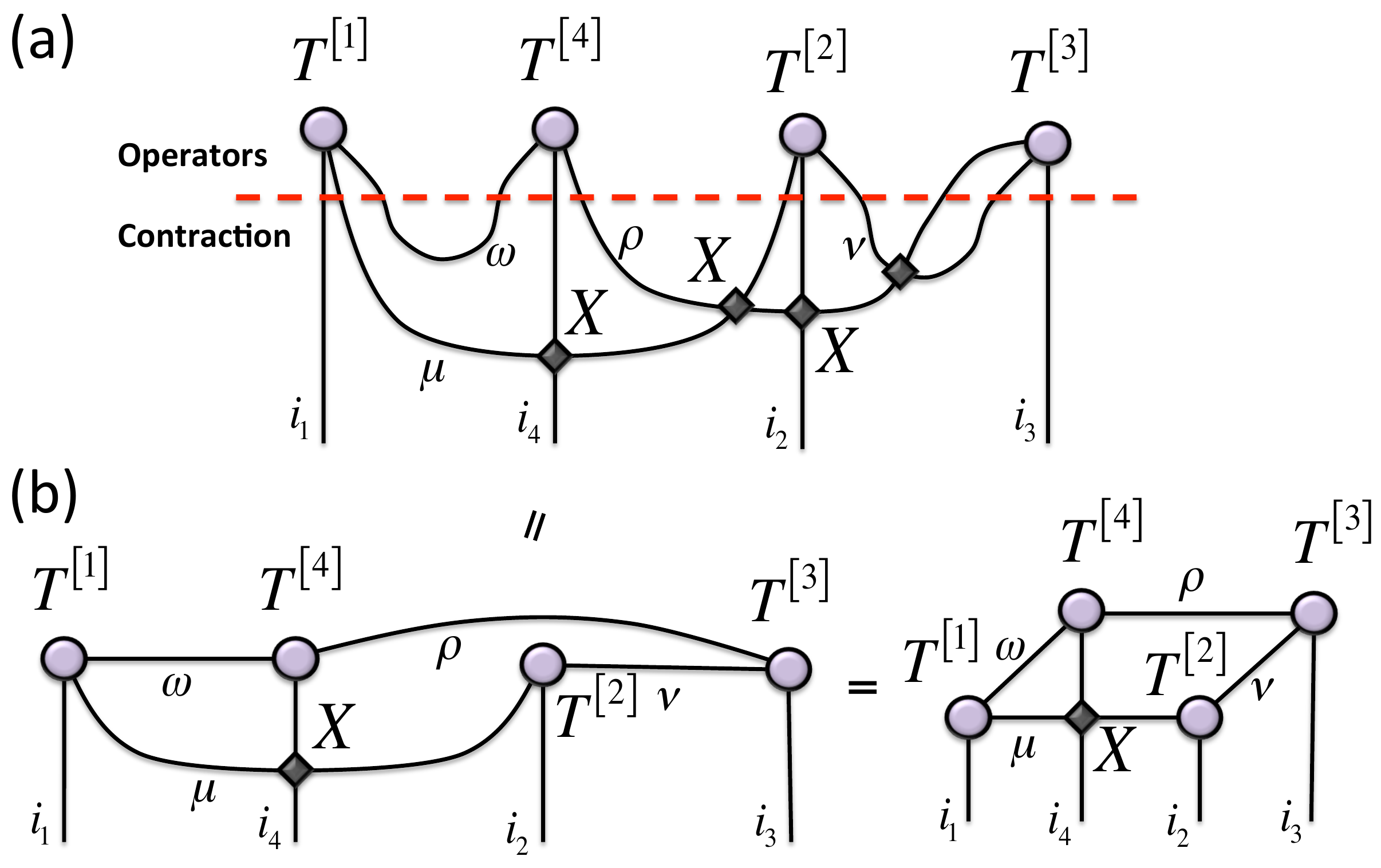} 
\par\end{centering}
\caption{(color online) Fermionic TN diagram for a 4-site MPS with periodic boundary conditions: the contraction of operators in (a) is equal to the diagrams in (b), since operator $T^{[2]}$ is parity-invariant and, therefore, index $\rho$ can \emph{jump} over it, avoiding several crossings and producing the usual diagram for a 4-site MPS (see main text).}
\label{Fig13}
\end{figure}

The above three examples illustrate the basic tricks of how a quantum state for fermions in second quantisation gets represented by a fermionic TN diagram using the two rules mentioned before. The reader is free to try more complicated examples and, using the basic ideas explained in the examples, arrive to the conclusion that the fermionisation rules stated at the beginning are in fact valid for any TN state. 

\subsection{Two properties}

\subsubsection{Fermion ordering = TN graphical projection}

From the examples above, as well as from further more complex structures, one realises the following important property: \emph{different fermionic orderings correspond to different graphical projections of the TN on a $2d$ plane}. More precisely, we see that different graphical projections of the TN on a plane (i.e., different ways of ``drawing'' it) correspond to different orderings of the physical fermionic modes. The minus signs needed to change the ordering by moving around the fermionic operators is taken into account, in the graphical projection, by the different fermionic swaps together with the parity symmetry of the tensors. Notice that this would also correspond to different orderings to implement a Jordan-Wigner transformation in order to map the fermionic system to a spin system. 

Importantly, notice that the above is an ``artificial'' degree of freedom. In the end, physical observables such as expectation values do not depend on the ordering that we choose for the physical fermions. Graphically this means that expectation values do not depend on the choice of graphical projection, even if the TN to be contracted has different fermionic swaps. In practice, two different projections can be mapped onto each other by jump moves such as the one used in the third example above, hence proving that observable quantities do not depend on the specific planar projection.

\subsubsection{No increase in the leading computational cost}

Another important property is that, using the fermionisation rules, the cost of simulating fermionic systems with TN methods has the same leading computational cost as the simulation of bosonic ones. Hence, fermionic statistics does not imply \emph{per se} an increase in the computational resources that are needed. For more details about this we refer the reader to Ref.\cite{fipeps}, where this is shown for $2d$ PEPS. In the end, one sees that all fermionic swaps can be accounted for locally in the $2d$ lattice, and the final $2d$ tensor network to be contracted is entirely bosonic (i.e., without fermionic swaps). Thus, the increase in computational cost for a given value of the bond dimensions is only sub-leading when changing from simulating bosons to simulating fermions. \color{black} Notice, however, that this does not mean that one should expect the same accuracy for a given bond dimension for bosons and fermions, since this mainly depends on the corresponding scaling of the entanglement entropy, which may be different. \color{black}

\color{black}
\subsection{Equivalence to other approaches}
The formalism reviewed in this paper based on graphical projections is equivalent to other existing approaches proposed to simulate fermionic systems with tensor networks states. We now briefly sketch why this is the case. 

First, let us remind the concept of fermionic PEPS (or fPEPS) as introduced in Ref.\cite{fPEPS}. Following the notation in that reference, for a square lattice at a given node $(h,v)$ four ancillary (spinless) fermionic modes are considered with creation operators $(\alpha^{[h,v]})^\dagger, (\beta^{[h,v]})^\dagger, (\gamma^{[h,v]})^\dagger, (\delta^{[h,v]})^\dagger$. One defines operators that create maximally entangled states of fermions from the vacuum: 
\beqa
H^{[h,v]}Ê&=& \frac{1}{\sqrt{2}}\left(1 + (\beta^{[h,v]})^\dagger (\alpha^{[h+1,v]})^\dagger \right) \nonumber \\
V^{[h,v]}Ê&=& \frac{1}{\sqrt{2}}\left(1 + (\delta^{[h,v]})^\dagger (\gamma^{[h,v+1]})^\dagger \right), 
\label{hv}
\eeqa
as well as "projectors" onto a physical (spinless) fermionic mode
\beq
Q^{[h,v]} = \sum T^{[h,v]}_{ilrud}(a^{[h,v]})^{\dagger i} (\alpha^{[h,v]})^l (\beta^{[h,v]})^r (\gamma^{[h,v]})^u (\delta^{[h,v]})^d , 
\eeq
with $a^{[h,v]}$ the anihilation operator of the physical fermion. The sum in the last expression runs for all indices from $0$ to $1$, with the parity constraint $(u + d + l + r + i){\rm mod} 2 = c$ with $c$ fixed for each node (which guarantees that the fPEPS has a well-defined parity). The fPEPS is then defined as
\beq
\ket{\Psi} = \left\langle \prod_{[h,v]} Q^{([h,v]} H^{[h,v]} V^{[h,v]} \right\rangle _{{\rm aux}} \ket{\Omega}, 
\eeq
where the expectation value is taken over the vacuum of the ancillary degrees of freedom. This way of defining fPEPS generalizes easily to the case of more than one ancillary mode per bond, physical fermions with inner structure (e.g., spin and color), as well as to other dimensions and network structures. 

Even if not obvious at first sight, the fPEPS constructed in this way is actually equivalent to the PEPS that one would get using the fermionic diagrammatic language described before. Long story short, one can see that (i) taking the expectation value over the ancillary fermionic degrees of freedom in the fPEPS representation is equivalent, in our language, to have "matching" virtual operators between different sites and a sum (trace) over their respective indices; (ii) tensors have a well-defined parity in both cases; and finally (iii) possible anticommutation factors in fPEPS are precisely those accounted by fermionic swap gates in the diagrammatic language. 

Let us sketch briefly why this is the case with the simple example of a 2-site MPS. For this, consider four fermionic virtual modes (2 per physical site) with anihilation operators $\alpha^{[1]}, \beta^{[1]}, \gamma^{[2]}$ and $\delta^{[2]}$ and two physical fermionic modes $a^{[1]}$ and $a^{[2]}$. Modes $\alpha^{[1]}$ and $\delta^{[2]}$ are "unpaired" edge modes, while we consider maximal entanglement between modes $\beta^{[1]}$ and $\gamma^{[2]}$. Following the fPEPS prescription from above, we get the fermioinc 2-site MPS
\beq
\ket{\Psi}Ê= \sum_{i_1 i_2 \mu} T^{[1]}_{i_1 \mu} T^{[2]}_{\mu i_2} \left\langle (a^{[1] \dagger})^{i_1} (\alpha^{[1]}) (\beta^{[1]})^\mu (a^{[2] \dagger})^{i_2} (\gamma^{[2]})^\mu (\delta^{[2]}) (\alpha^{[1]})^\dagger (\delta^{[2]})^\dagger (1+(\beta^{[1]})^\dagger (\gamma^{[2]})^\dagger) \right\rangle_{{\rm aux}}  \ket{\Omega},   
\eeq
where unpaired edge modes are unentangled. After taking the expectation value over the ancillary fermionic modes, and using fermionic anticonmutation relations, the above equation can be reduced to 
\beq
\ket{\Psi} = \sum_{i_1  i_2} \left( \sum_\mu T^{[1]}_{i_1 \mu} T^{[2]}_{\mu i_2}\right)  (a^{[1] \dagger})^{i_1}  (a^{[2] \dagger})^{i_2} \ket{\Omega}, 
\eeq
which is nothing but Eq.\ref{2mp} for spinless physical (and virtual) fermions. 

The above example illustrates how the two formalisms are equivalent. Thus, calculations in one setting are equivalent to calculations in the other. Indeed, several works have proposed schemes to contract and manipulate the fermionic anticommutation signs present in fPEPS \cite{pit}, including the so-called fermionic operator circuits \cite{foc}. Because of the mentioned equivalence these schemes must correspond, in the end, to specific choices of graphical projections in the language of fermionic tensor network diagrams, giving rise to specific patterns of fermionic swap gates. 

\color{black} 

\subsection{Recent implementations}

Fermionic systems in $2d$ where initially addressed in Refs.\cite{fmera, fipeps, fPEPS, pit, foc}. In Ref.\cite{fmera}, the fermionic $2d$ MERA was presented, and in Ref.\cite{fipeps} the present graphical language was developed in the context of the iPEPS algorithm for $2d$ systems \cite{iPEPS}. Fermionic iPEPS has been later used to simulate a variety of spinless \cite{fspinless} and spin-full models, including some promising results for the $t - J$ model \cite{ftj}. Fermionic PEPS have also been recently used in the context of resonating valence-bond (RVB) wavefunctions \cite{frvb}.  

\section{TNs and entanglement spectrum: Hamiltonians}

The study of entanglement in many-body systems is by now a well established field of research \cite{arealaw, 2darea, revent}. By considering different measures of entanglement, people have shown how different regimes of quantum matter are interesting and important. Within this setting, a key development was due to Li and Haldane in Ref.\cite{liha}, who pointed out that the eigenvalues of the reduced density matrix of a bipartition codify, in a holographic way, information about the boundaries of the system. Somehow, when a many-body system is partitioned into two pieces, the boundary of the bipartition acts as a ``sort of'' an artificial boundary for the system, so that boundary physics can be read from the entanglement properties. To be more specific, in Ref.\cite{liha} they showed that the spectrum of eigenvalues of the reduced density matrix of a subsystem (known as \emph{entanglement spectrum}) for the fractional quantum Hall effect wave-function was in one-to-one correspondence to the energy spectrum of a conformal field theory describing the degrees of freedom at the boundary. In fact, the reduced density matrix of a bipartition can be written as 
\beq
\rho \propto e^{- H_E}, 
\eeq
where $H_E$ is the so-called \emph{entanglement Hamiltonian}. The claim is that such entanglement Hamiltonan describes, at low energies, the fundamental degrees of freedom of the projection of the quantum state on a boundary, hence being this a natural realisation of the holographic principle in terms of entanglement. 

It turns out that TNs are a natural arena to investigate entanglement Hamiltonians for a variety of systems, very specially for $2d$ PEPS. In what follows we sketch briefly how this is done, together with some important results along these lines. The interested reader is addressed to Ref.\cite{enthampeps} for more details. 

\subsection{Entanglement Hamiltonians from $2d$ PEPS}

\begin{figure}
\begin{centering}
\includegraphics[width=8cm]{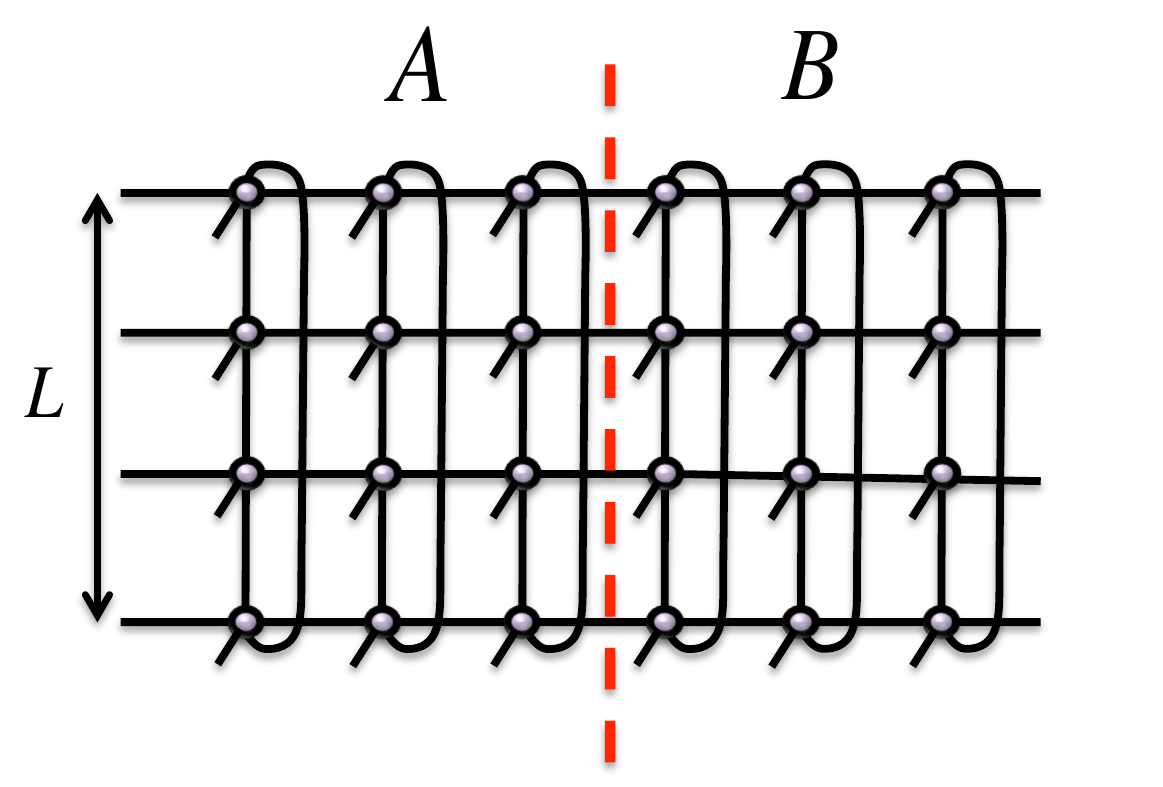} 
\par\end{centering}
\caption{(color online) $2d$ PEPS on a cylinder, partitioned in two pieces $A$ and $B$.}
\label{Fig16}
\end{figure}

Consider a $2d$ PEPS $\ket{\Psi}$ wrapped around a cylinder of circumference $L$, as in Fig.\ref{Fig16}. The cylinder can either be finite, as in Ref.\cite{enthampeps}, or infinite, as in Ref.\cite{sharp}.
As explained in Ref.\cite{enthampeps}, the reduced density matrix of half an infinite cylinder (e.g., $A$) is given by
\beq
\rho = U \sqrt{\sigma_A^T}\sigma_B \sqrt{\sigma_A^T} U^\dagger,
\eeq
with $\sigma_{A/B}$ the reduced density operators in $A/B$ for the virtual spaces across the bipartition, and $U$ an isometry given by the contraction of the PEPS tensors. For simplicity of this explanation, let us assume that appropriate space inversion symmetries are present in the PEPS (this could be generalised easily to other scenarios \cite{enthampeps}), so that $\sigma_A = \sigma_B \equiv \sigma$. Then we have that 
\beq
\rho = U \sigma^2 U^\dagger.
\eeq
The above equation implies that $\rho$ and $\sigma^2$ are isospectral, since the isometries do not change the eigenvalue spectrum.  Therefore, we can say that 
\beq
\sigma^2 \propto e^{- H'_E}. 
\eeq
As a result we have that $H_E = U H'_E U^\dagger$, so both Hamiltonians are isospectral. The isometry only amounts to a change of basis in the eigenvectors. Since the spectrum of eigenvalues of a Hamiltonian is dictated by the type of its interactions, it thus makes sense to study what types of interacting Hamiltonians $H'_E$ does one get when performing such a calculation for different types of $2d$ PEPS. Notice also that, in this way, Hamiltonians emerge naturally from entanglement via the holographic projection of a TN state on a ``fake'' boundary. 

\subsection{Holographic principle at work}

Given the derivations above, one can start considering different kinds of $2d$ PEPS states and see what types of entanglement Hamiltonians one gets. So far there are no generic theorems and all evidence is based on a plethora of examples, see Refs.\cite{enthampeps, ehexamples}. Quite interestingly, it looks like there is an emergent picture from the examples analysed so far. To say it in few words, consider PEPS that are ground states of $2d$ Hamiltonians with local interactions. If the $2d$ system is gapped and not topologically ordered, then it looks like $H'_E$ is usually a $1d$ Hamiltonian with short-range interactions. If the $2d$ system is critical, then $H'_E$ is a $1d$ Hamiltonian with long-range interactions (i.e., the interaction length in the system becomes very large). Finally, if the $2d$ system is gapped and topologically ordered, then $H'_E$ is completely non-local, e.g., close or equal to a projector.

The above picture is very appealing. Even if based on examples, it is nothing but the holographic principle at work,  giving a nice example of a bulk-boundary correspondence at the level of $2d$ PEPS. Moreover, this picture justifies something that people have been finding for quite some time in numerical simulations with $2d$ PEPS, namely, that environment calculations in, e.g., infinite $2d$ PEPS converge quickly with very few iterations of a boundary MPS \cite{PEPS2, iPEPS, fipeps}. Indeed, for a $2d$ gapped system, if the boundary entanglement Hamiltonian is $1d$ and short range, then its thermal state can always be approximated with very good accuracy by a matrix product operator \cite{Hastings}. Hence it may happen that, in the end, $2d$ PEPS corresponding to gapped $2d$ systems with no topological order can actually be contracted efficiently with good accuracy, even if this is not true for arbitrary PEPS \cite{sphard}. Again this is only based on specific examples and numerical observations, and a more precise theorem along these lines would be very useful. 

\section{TNs and gauge/gravity duality: geometry}

In recent years, another inspiring connection has been noticed between entanglement,  TNs, and quantum gravity. This is a relation between Vidal's MERA and Maldacena's AdS/CFT or gauge / gravity duality \cite{adscft}. Such a connection is opening new intriguing research directions, and advocates the point of view that geometry emerges from the structure of entanglement in quantum many-body states. The key relation between these two concepts was explicitly noticed first by Swingle in Ref.\cite{Swingle1}, and subsequently  developed in several further contributions by many authors \cite{Swingle2, Ryu, adsmera}. In fact, the connection is also reinforced by field-theory results on the calculation of entanglement entropies via holography by Ryu and Takayanagi \cite{ryutaka}, which perfectly matches the MERA results. In what follows we provide a quick overview on the basics of this connection, in a rather non-technical way.  The reader interested in technical details and specific developments is referred to Refs.\cite{Swingle1, Swingle2, Ryu, adsmera}. 

\subsection{From AdS/CFT to AdS/MERA}

Maldacena proposed a conjecture, later refined by Gubser, Klebanov, Polyakov and Witten \cite{adscft}, that changed our understanding of quantum gravity and quantum field theories. This was the holographic gauge/gravity duality, or AdS/CFT correspondence, which states that some quantum field theories without gravity are dual to quantum gravity theories in a curved higher-dimensional ``bulk'' geometry. Holographic projections provide prescriptions to compute observables in field theory from a purely geometrical point of view in the corresponding gravitational dual. In particular, conformal field theories (CFT) have a gravity dual in an Anti-de Sitter space (AdS), which is a space with negative curvature. The duality has been applied very successfully not only in quantum gravity, but also in the context of strongly-correlated systems \cite{adsconmat}. 

\begin{figure}
\begin{centering}
\includegraphics[width=12cm]{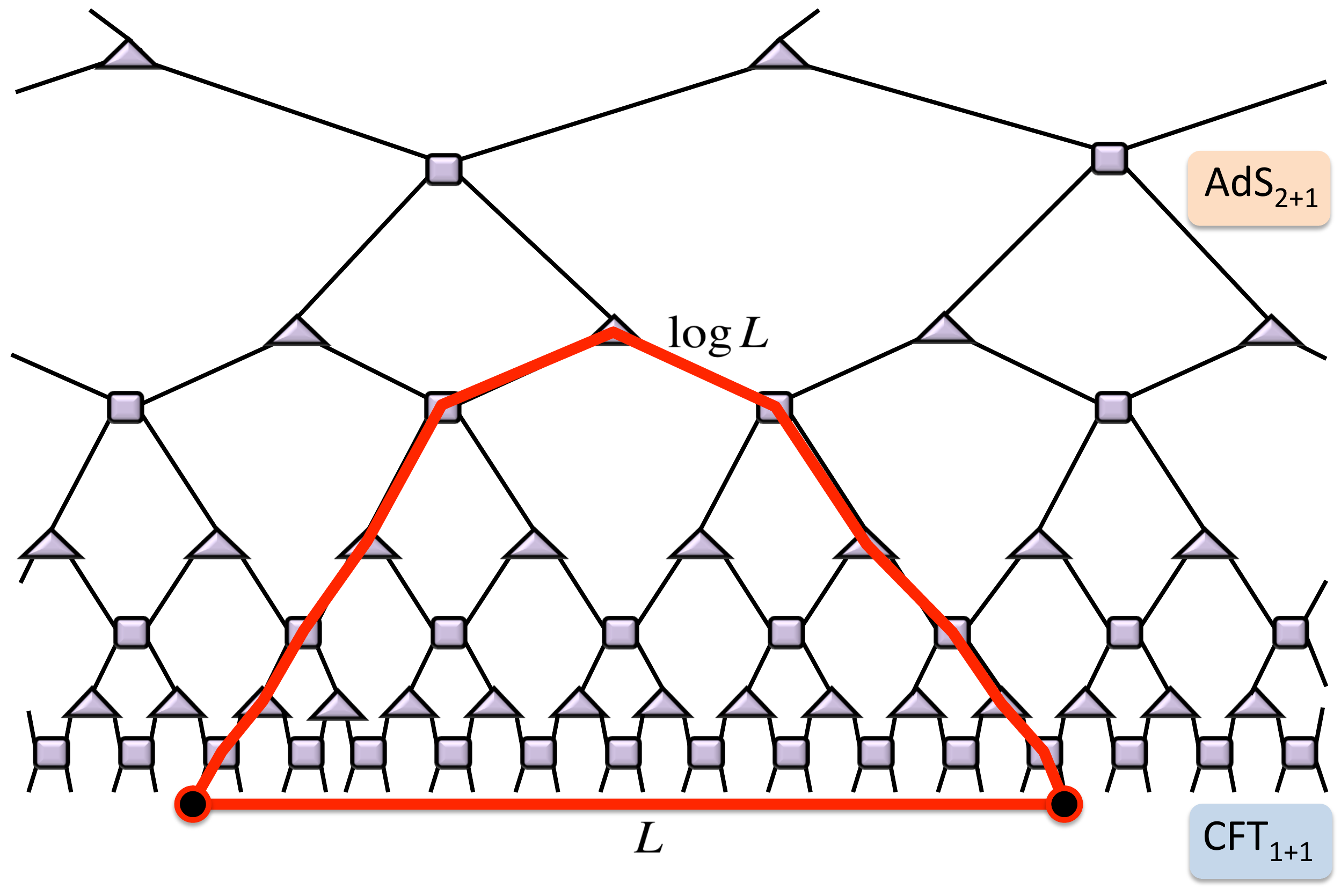} 
\par\end{centering}
\caption{(color online) In scale-invariant $1d$ MERA, two sites separated by distance $L$ in physical space are separated by distance $O(\log L)$ in holographic space. If the MERA reproduces the properties of a $1d$ quantum critical lattice system with a CFT limit, then one can understand the open indices in the TN as a discrete version of a $(1+1)d$-CFT, living at the boundary of a bulk that is  a discrete version of a $(2+1)d$-AdS gravitational dual.}
\label{Fig18}
\end{figure}

Let us now come back to TNs and consider, for simplicity, a critical $1d$ quantum lattice system. This could be, for instance, the spin-$1/2$ quantum Ising model in $1d$ and with a transverse magnetic field at its critical value. It is well known that such a quantum lattice model admits a field theory description in terms of a massless fermionic quantum field theory \cite{arealaw}, which has conformal symmetry and, hence, is a $(1+1)$-dimensional CFT with central charge $c = 1/2$. Forgetting about the field-theory description, one could try to reproduce the properties of the ground state of this quantum lattice system by means of a scale-invariant MERA such as the one in Fig.\ref{Fig18}. This TN ansatz has in fact the correct entanglement structure in order to host a logarithmic scaling of the entropy of a block as well as power-law decaying correlation functions, as happens for this critical $1d$ spin system. Thus, the appropriate MERA tensors will produce a TN representation of the corresponding quantum ground state of this system. 

Notice now that we can understand the MERA in terms of a ``bulk'' of tensors that span along one physical and one holographic (renormalisation) dimension. The bulk of tensors defines a geometry, in the sense that one can measure distances, say, between two given tensors, by counting the minimum (i.e., geodesic) number of jumps in the TN needed to go from one tensor to another. Understood in this way, one arrives at the conclusion that the bulk of tensors of a scale-invariant MERA spans a discretised version of a space with negative curvature, i.e., an AdS geometry \cite{tngeo}. See Fig.\ref{Fig18} for a graphical explanation. 

Now put the concepts from the previous two paragraphs together: for a scale-invariant MERA, the tensors in the bulk can be understood as a discretised AdS geometry \emph{and} the physical degrees of freedom at the boundary as a discretisation of a CFT. Thus, what we have is nothing but a \emph{lattice realisation of the AdS/CFT correspondence}. This is what some people call the AdS/MERA duality \cite{adsmera}. Such a connection can be formalised in different ways, e.g., by taking the continuum limit of the MERA \cite{cMERA} and evaluating the metric of the resulting smooth space \cite{Swingle2, Ryu}. 

Let us now think for a second a bit more about the above connection. We explained before that the construction of the MERA is motivated by the pattern of entanglement in quantum critical states of matter. In these states entanglement is built locally at every length scale, and this produces naturally a MERA TN. Now we see that, in fact, for a quantum critical system the bulk of tensors that is spanned by the MERA can be interpreted as some sort of gravitational dual of the quantum theory living in the \emph{open indices} at the boundary. When seen from this perspective, one would say that geometry (and gravity) seems to emerge from local patterns of entanglement in quantum many-body states. 

Moreover, the connection opens fruitful lines of research along several directions. First, known results in AdS/CFT could now inspire new \emph{ans\"atze}  for complicated quantum many-body wave-functions, just by assuming that gravitational theories may correspond to given patterns of entanglement yet to be explored \cite{adsresults}. Second, results in TN methods and quantum information theory may now find an even more direct application in quantum gravity and quantum field theory, e.g., in the study of black holes. In fact, judging by the connection between AdS/CFT and the MERA, it would not be surprising if the language of TNs provided further insight on the quantum structure of space-time at small length scales. After all, other results coming from different directions seem to support further the idea that \emph{gravitational space-time emerges from quantum entanglement}. Two examples of this are the space-time constructions by Van Raamsdonk \cite{vR}, and the recent $ER = EPR$ conjecture by Maldacena and Susskind \cite{erepr} in relation to the black hole firewall paradox, and which is sustained already by a number of examples \cite{ereprexamples}. \color{black} This line of research is extremely suggestive, but not yet theoretically established at the level of other developments. \color{black}

\section{Final remarks: towards a ``quantum gravity" era of TNs?}

Here we did a quick review of some important theoretical aspects of TN states for strongly-correlated systems. We focused on symmetries, fermions, entanglement Hamiltonians, and the connection to the gauge/gravity duality. These developments have many implications, both theoretical and numerical. Of course, we did not explain here many other interesting developments in TN theory. But we decided to focus on these four since we believe that, when seen in perspective, they offer a very nice picture of the theory of TN expanding along different scientific directions, some of them quite unforeseen just a few years ago. 

Other recent developments not considered here, but which are also of importance, include the relation between symmetries and  string/membrane order in TNs \cite{injpeps, david}, parent and uncle Hamiltonians \cite{parent, uncle}, $1d$ symmetry-protected topological order from entanglement spectrum in MPS \cite{pollmann}, anyonic systems \cite{anyonsTN}, the classification of quantum phases of matter \cite{phases}, $2d$ PEPS with chiral topological order \cite{chiralPEPS}, continuous TNs \cite{cMPS, cMERA, cTN}, TN representations of the Bethe ansatz \cite{tnbethe} and the fermionic Fourier transform \cite{tnfft}, Monte Carlo sampling of TNs \cite{mctn}, and single-layer numerical algorithms \cite{sltn}, to name a few. 

Looking in perspective, the historical development of TN methods has followed different periods. One could talk of an initial ``statistical mechanics'' era underpinned by results in exactly solvable models (specially by Baxter \cite{baxter}) around the 60's and 70's. After that period we could say that a ``DMRG era'' started in the 90's with the explosion of White's DMRG for $1d$ systems and subsequent applications. Later on, around 2000's one could talk about the advent of a ``quantum information era'' with the many results on many-body entanglement and further TN developments. As for today, one could perhaps talk about a new ``quantum gravity era" of TNs that is just starting. In fact, it looks like important physical objects, such as curved space-times and quantum Hamiltonians, emerge naturally from entanglement in TN states via holography.  The future looks certainly very exciting. 

\noindent {{\bf Acknowledgements}}
The author acknowledges financial support from the Johannes Gutenberg-Universit\"at, as well as the participants of the \emph{CMSI International Workshop 2013} in Kobe, the \emph{DPG Spring Meeting 2014 of the Condensed Matter Section} in Dresden, the \emph{School on computational methods in quantum materials} in Jouvence, and the \emph{4th Les Houches School in Computation Physics} in Les Houches, for some very interesting discussions which partly motivated this paper. Insightful comments by M. Rizzi and P. Corboz are also acknowledged.  

{}

\end{document}